\documentstyle[preprint,eqsecnum,aps]{revtex}

\def\lesssim{\ \raise.3ex\hbox{$<$}\kern-0.8em\lower.7ex\hbox{$\sim$}\ }
\def\gesim{\ \raise.3ex\hbox{$>$}\kern-0.8em\lower.7ex\hbox{$\sim$}\ }

\font\scripti=cmmi7
\font\scriptscripti=cmmi5
\def\sib#1{\setbox0 = \hbox{\scripti #1}
  \kern-.02em\copy0\kern-\wd0
  \kern.04em\box0} 
\def\ssib#1{\setbox0 = \hbox{\scriptscripti #1}
  \kern-.02em\copy0\kern-\wd0
  \kern.04em\box0} 

\font\tenib=cmmib10 
\skewchar\tenib='177 \skewchar\tenib='177 \skewchar\tenib='177
\def\pbold#1{\setbox0 = \hbox{$ #1 $}
  \kern-.022em\copy0\kern-\wd0
  \kern.011em\copy0\kern-\wd0
  \kern.011em\copy0\kern-\wd0
  \kern.011em\copy0\kern-\wd0
  \kern.011em\box0} 

\begin{document}
\draft
\title{Superfluid transition temperature in a trapped gas of Fermi atoms with a Feshbach resonance}
\author{Y. Ohashi$^{1,2}$ and A. Griffin$^1$}
\address{$^1$Department of Physics, University of Toronto, Toronto, Ontario, Canada M5S 1A7\\
$^2$Institute of Physics, University of Tsukuba, Ibaraki 305, Japan}
\maketitle
\begin{abstract}
We investigate strong coupling effects on the superfluid phase transition in a gas of Fermi atoms with a Feshbach resonance. The Feshbach resonance describes a composite quasi-Boson, which can give rise to an additional pairing interaction between the Fermi atoms. This attractive interaction becomes stronger as the threshold energy ($2\nu$) of the Feshbach resonance two-particle bound state is lowered. In a recent paper, we showed that in the uniform Fermi gas, this tunable pairing interaction naturally leads to a BCS-BEC crossover of the Nozi\`eres and Schmitt-Rink kind, in which the BCS-type superfluid phase transition continuously changes into the BEC-type as the threshold energy is decreased. In this paper, we extend our previous work by including the effect of a harmonic trap potential, treated within the local density approximation (LDA). We also give results for both weak and strong coupling to the Feshbach resonance. We show that the BCS-BEC crossover phenomenon strongly modifies the shape of the atomic density profile at the superfluid phase transition temperature $T_{\rm c}$, reflecting the change of the dominant particles going from Fermi atoms to composite Bosons. In the BEC regime, these composite Bosons are shown to first appear well above $T_{\rm c}$. We also discuss the ``phase diagram" above $T_{\rm c}$ as a function of the tunable threshold energy $2\nu$. We introduce a characteristic temperature $T^*(2\nu)$ describing the effective crossover in the normal phase from a Fermi gas of atoms to a gas of stable molecules.
\par
\vskip3mm
\end{abstract}
\pacs{PACS numbers: 05.30.Jp, 03.75.Fi,67.90.+z}
\narrowtext
%
\section{Introduction}
The search for the superfluid phase transition in a trapped gas of Fermi atoms is one of the most exciting problems in current physics.\cite{Stoof} This involves the formation of BCS Cooper-pairs made up of degenerate Fermions of two different species (``spin up" and ``spin down"). Degenerate Fermi gases have been realized in $^{40}$K\cite{Jin} and $^6$Li.\cite{Andrew,Salomon,Granade} Very recently, a Feshbach resonance was observed in $^{40}$K working with the two hyperfine states $|9/2,-9/2\rangle$ and $|9/2,-7/2\rangle$.\cite{Loftus} In the Feshbach resonance, a quasi-molecular Boson is formed, which can produce an attractive pairing interaction between two Fermi atoms of opposite spins.\cite{Timmermans1,Timmermans2,Holland,Chiofalo,Ohashi} As a result, one expects a BCS-type superfluidity with a high phase transition temperature $T_{\rm c}$ due to this strong attractive interaction.\cite{Holland} However, such strong interaction is also known to enhance fluctuations, which can strongly suppress $T_{\rm c}$ predicted by the usual mean-field BCS theory. Thus the inclusion of these Cooper-pair channel fluctuations is crucial in considering the high-$T_{\rm c}$ superfluidity induced by an atomic Feshbach resonance.
\par
In a previous paper,\cite{Ohashi} we have discussed how this fluctuation contribution affects the superfluid phase transition temperature in a uniform Fermi gas, extending the theory developed by Nozi\`eres and Schmitt-Rink:\cite{Nozieres,Tokumitsu,Melo,Randeria} We assume that $N_\uparrow=N_\downarrow$, where $N_\sigma$ is the number of Fermi atoms with spin-$\sigma$, and that both Fermion components have the same single-particle energy. As the threshold energy of the Feshbach resonance is lowered, the character of the phase transition is found to continuously change from the BCS-type to BEC-type. In the latter regime, $T_{\rm c}$ is strongly suppressed compared with the usual mean-field BCS theory and it approaches $T_{\rm c}=0.218T_{\rm F}$ in the BEC limit. (Here $T_{\rm F}$ is the Fermi temperature of the non-interacting spin-up Fermi gas.) We show that the stable Bosons which appear as we enter the BEC regime are a strongly hybridized mixture of Feshbach molecules and preformed Cooper-pairs. The Feshbach component of these Bosons dominates the excitation spectrum in the BEC limit.
\par
In this paper, we extend our previous work\cite{Ohashi} by including the effect of a harmonic trap potential. This extension is clearly necessary since all experiments on Fermi gases are done in some sort of trap potential. So far, the trapping potential has been examined within the mean-field BCS theory using the simple local density approximation (LDA).\cite{Chiofalo} In this paper, we also use the LDA but go past the mean-field approximation to include the effect of particle-particle fluctuations. We show how the BCS-BEC crossover can be observed from characteristic changes in the static atomic density profile, easily measured by standard techniques. 
\par
Besides the phase transition temperature $T_{\rm c}$, it is also important to understand how strong-coupling fluctuation effects enter above $T_{\rm c}$. In particular, it is an interesting problem to clarify at what temperature the stable Feshbach molecules and preformed Cooper-pairs in the BEC regime first appear in the normal-fluid phase above $T_{\rm c}$. Our calculations show that above $T_{\rm c}$, there exists a characteristic temperature $T^*$ which describes the crossover from a free Fermi gas to a molecular Bose gas. 
\par
In this paper, our treatment of the particle-particle fluctuations assumes that their coupling to the Feshbach resonance is weak and can be thus treated perturbatively. Most of our discussion is for this case, in which the BCS-BEC crossover is driven by decreasing the value of the Feshbach resonance threshold ($2\nu$) relative to twice the bare Fermi energy ($2\varepsilon_{\rm F}$). When the Feshbach coupling strength is very large, as Kokkelmans and coworkers have found,\cite{Kokkelman} it is not clear that our simple treatment of the particle-particle fluctuations is valid. However, in this limit, one finds that the BCS-BEC crossover occurs for values $2\nu\gg2\varepsilon_{\rm F}$ because of the broad Feshbach resonance (see also Ref.\cite{Milstein}). Effectively, one is in the strong-coupling regime discussed by Nozi\`eres and Schmitt-Rink,\cite{Nozieres} where the BEC phase is associated with preformed Cooper-pairs. When the threshold energy $2\nu$ becomes comparable to $2\varepsilon_{\rm F}$, our calculations again show that the Feshbach resonance takes over as the dominant component of the BEC phase.
\par
This paper is organized as follows: In Section II, we review the coupled Fermion-Boson model used in earlier work and discuss the LDA extension of the results we presented in Ref.\cite{Ohashi}. In Section III, we use this opportunity to expand on certain aspects which were not discussed in detail in Ref. \cite{Ohashi} due to lack of space. The BCS-BEC crossover in a trapped system is discussed in Section IV. In Section V, we consider the character of the composite Bosons. The crossover from a free Fermi gas to a gas of long-lived molecular Bosons above $T_{\rm c}$ is discussed in Section VI. Section VII considers the case of a strong Feshbach coupling. 
\par
\vskip3mm
\section{Formulation}
\subsection{Coupled Fermion-Boson gas in a trap}
The gas of interacting Fermi atoms coupled to a Feshbach resonance can be described by the coupled Fermion-Boson model given by\cite{Timmermans1,Timmermans2,Holland,Chiofalo,Ohashi,Lee,Ranninger1,Ranninger2} 
\begin{eqnarray}
{\cal H}
&=&
\sum_{{\sib p}\sigma}
\varepsilon_{\sib p}
c_{{\sib p}\sigma}^\dagger c_{{\sib p}\sigma}
+
\sum_{\sib q}(E^0_{\sib q}+2\nu)b_{\sib q}^\dagger b_{\sib q}
\nonumber
\\
&-&
U
\sum_{{\sib p},{\sib p}'}
c_{{\sib p}\uparrow}^\dagger
c_{-{\sib p}\downarrow}^\dagger
c_{-{\sib p}'\downarrow}
c_{{\sib p}'\uparrow}
+
g_{\rm r}
\sum_{{\sib p},{\sib q}}
[
b_{\sib q}^\dagger
c_{-{\sib p}+{\sib q}/2\downarrow}
c_{{\sib p}+{\sib q}/2\uparrow}
+{\rm h.c}
].
\label{eq.2.1}
\end{eqnarray}
Here $c_{{\sib p}\sigma}^\dagger$ is the creation operator of a Fermi atom with the kinetic energy $\varepsilon_{\sib p}=p^2/2m$. We assume that the system involves two atomic hyper-fine states which will lead to Cooper-pair Bosons and hence superfluidity. We describe these atomic states for simplicity using the pseudo-spin language $\sigma=\uparrow,\downarrow$. The quasi-molecular Boson of momentum ${\bf q}$ associated with the Feshbach resonance is described by the operator $b_{\sib q}^\dagger$. $E^0_{\sib q}\equiv q^2/2M$ is the kinetic energy of this Boson and $2\nu$ represents the bottom of this resonance Bose spectrum, to be referred to as the threshold energy of the Feshbach resonance. The coupling to this Feshbach resonance Boson is described by last term in eq. (\ref{eq.2.1}) with a strength $g_{\rm r}$, in which two Fermi atoms form one $b$-Boson and the $b$-Boson breaks into two free Fermi atoms. The Hamiltonian also includes a standard BCS pairing interaction with the coupling constant $-U<0$, originating from non-resonant interactions between two Fermi atoms ($\sigma=\uparrow,\downarrow$). Even assuming that this direct attractive pairing interaction $U$ is weak, the effective pairing interaction strength can be strong due to the effect of the coupling to the Feshbach resonance. In this paper, we do not discuss the question of how good the model Hamiltonian in eq. (\ref{eq.2.1}) is in the context of ultracold Fermi gases, but simply refer to the literature \cite{Timmermans1,Timmermans2,Holland,Chiofalo,Kokkelman}.
\par
Since in this model, one Feshbach-induced $b$-molecule consists of two Fermi atoms, we have $M=2m$ and also impose the conservation of the total number of particles as 
\begin{eqnarray}
N
&=&
\langle \sum_{{\sib p}\sigma}c_{{\sib p}\sigma}^\dagger c_{{\sib p}\sigma}\rangle+2\langle\sum_{\sib q}b_{\sib q}^\dagger b_{\sib q}\rangle
\nonumber
\\
&\equiv&
N_{\rm F}+2N_{\rm B}.
\label{eq.AGY}
\end{eqnarray}
As discussed in our previous paper,\cite{Ohashi} this constraint is incorporated into the model Hamiltonian in eq. (\ref{eq.2.1}) by replacing ${\cal H}\to{\cal H}-\mu N$, where $\mu$ is the chemical potential. The resulting Hamiltonian has the same form as eq. (\ref{eq.2.1}), but the kinetic energies of the Fermions and Bosons are now given by $\varepsilon_{\sib p}\to\varepsilon_{\sib p}-\mu$ and $E^0_{\sib q}\to E^0_{\sib q}-2\mu$, respectively. 
\par
Now we introduce the effect of a harmonic trap. In this paper, we consider an optical trap in which the two hyperfine Fermi atom states feel the same trap potential given by 
\begin{equation}
V_{\rm trap}({\bf r})
={1 \over 2}\sum_jm
[
\omega_{0x}^2x_j^2+ 
\omega_{0y}^2y_j^2+ 
\omega_{0z}^2z_j^2 
],
\label{potential}
\end{equation}
where ${\bf r}_j=(x_j,y_j,z_j)$ is the position of the $j$-th atom. In our model, we also assume that the Feshbach molecules also feel this potential with $m$ being replaced by the Boson mass $M=2m$. We could relax this last assumption to deal more realistically with the dipole moment of the composite molecule but we leave this extension to future studies. Treating the resulting trapped Fermion-Boson gas within the LDA\cite{Molmer,Minguzzi} amounts to replacing the chemical potential $\mu$ by
\begin{equation}
\mu({\bf r})=\mu-V_{\rm trap}({\bf r}).
\label{potential2}
\end{equation}
The constraint on the total number of Fermi atoms is then replaced by $N=\int d {\bf r}[N_{\rm F}({\bf r})+2N_{\rm B}({\bf r})]$, where $N_{\rm F}({\bf r})$ and $N_{\rm B}({\bf r})$ are the number density of Fermi atoms and Bose molecules, respectively.
\par
The atomic hyperfine states may feel different trap potentials in a magnetic trap. This situation is similar to electron spins in superconducting metals under a magnetic field. In the theory of superconductivity, it is known that a magnetic field suppresses superconductivity and, under certain conditions, produces a non-uniform kind of superconducting state called the Fulde-Ferrell state.\cite{Fulde,Larkin,Takada} The effect of fluctuations on this kind of Zeeman splitting in atomic Fermi gases is considered in another paper.\cite{OhashiFF}
\par
\vskip3mm
\subsection{Equations for $T_{\rm c}$: Thouless criterion}
\vskip2mm
The superfluid phase transition temperature $T_{\rm c}$ can be conveniently determined using the Thouless criterion,\cite{Nozieres} in which the phase transition is characterized by a singularity in the particle-particle scattering vertex function $\Gamma$ at $\omega={\bf q}=0$, associated with the formation of Cooper-pairs.\cite{Baym,Kadanoff} The four-point vertex function $\Gamma$ within the generalized $t$-matrix approximation in terms of $U$ and $g_{\rm r}$ is given by the diagrams in Fig.1. In this figure, the first line describes ladder processes associated with the attractive interaction $-U<0$, processes familiar in strong coupling superconductivity in electronic systems.\cite{Nozieres} On the other hand, the second line includes the effective Fermion-Fermion interaction mediated by the $b$-Boson propagator $D_0$ defined in eq. (\ref{eq.2.3}), scattering processes which only arise in atomic Fermi gases coupled to a Feshbach resonance, such as in eq. (\ref{eq.2.1}). The equation of the vertex function $\Gamma$ is given by
\begin{eqnarray}
&\Gamma&(p_+,-p_-;p'_+,-p'_-;{\bf r})
=
-(U-g_{\rm r}^2D_0(q;{\bf r}))
\nonumber
\\
&+&
(U-g_{\rm r}^2D_0(q;{\bf r})){1 \over \beta}\sum_{{\sib p}'',\omega_m''}
G_0(p''_+;{\bf r})G(-p''_-;{\bf r})
\Gamma(p''_+,-p''_-;p'_+,-p'_-;{\bf r}),
\label{eq.2.2}
\end{eqnarray}
where $p_+\equiv ({\bf p}+{\bf q}/2,i\omega_m+i\nu_n)$, $p_-\equiv ({\bf p}-{\bf q}/2,i\omega_m)$, $q\equiv ({\bf q},i\nu_m)$ and $\beta\equiv 1/T$ is the inverse of temperature; $i\omega_m$ and $i\nu_n$ are the Fermion and Boson Matsubara frequencies, respectively. $G_0$ and $D_0$ represent the bare one-particle Fermion and Boson thermal Green functions given by
\begin{eqnarray}
\left\{
\begin{array}{l}
\displaystyle
G_0({\bf p},i\omega_m;{\bf r})=
{1 \over i\omega_m-\varepsilon_{\sib p}+\mu({\bf r})},\\
\displaystyle
D_0({\bf q},i\nu_n;{\bf r})=
{1 \over i\nu_n-E^0_{\sib q}-2\nu+2\mu({\bf r})}.\\
\end{array}
\right.
\label{eq.2.3}
\end{eqnarray}
Since the modified chemical potential $\mu({\bf r})$ includes the trap potential, the LDA Green functions and vertex function $\Gamma$ also depend on position ${\bf r}$. Multiplying both the sides of eq. (\ref{eq.2.2}) by ${1 \over \beta}G_0(p_+;{\bf r})G_0(-p_-;{\bf r})$ and summing over $\omega_m$ and ${\bf p}$, we obtain
\begin{equation}
{1 \over \beta}\sum_{{\sib p}'',\omega_m''}
G_0(p''_+;{\bf r})G(-p''_-;{\bf r})
\Gamma(p''_+,-p''_-;p'_+,-p'_-;{\bf r})=
-{(U-g_{\rm r}^2D_0(q;{\bf r}))\Pi(q;{\bf r}) \over 1-(U-g_{\rm r}^2D_0(q;{\bf r}))\Pi(q;{\bf r})}.
\label{eq.2.3b}
\end{equation}
Here $\Pi(q;{\bf r})=\Pi({\bf q},i\nu_n;{\bf r})$ is the well-known particle-particle correlation function of the Cooper-pair-field operator defined by
${\hat \Delta}({\bf q},\tau;{\bf r})
\equiv\sum_{\sib p}c_{-{\sib p}+{\sib q}/2\downarrow}(\tau;{\bf r})
c_{{\sib p}+{\sib q}/2\uparrow}(\tau;{\bf r})$ in the absence of $U$ and $g_{\rm r}$:\cite{Ohashi,Nozieres,OhashiT,Takada2}
\begin{eqnarray}
\Pi({\bf q},i\nu_n;{\bf r})
&=&
\int_0^\beta d\tau e^{i\nu_n\tau}
\langle
T_\tau
\{
{\hat \Delta}({\bf q},\tau;{\bf r})
{\hat \Delta}^\dagger({\bf q},0;{\bf r})
\}
\rangle
\nonumber
\\
&=&{1 \over \beta}\sum_{{\sib p},\omega_m}
G_0({\bf p}+{\bf q}/2,i\omega_m+i\nu_n;{\bf r})
G_0(-{\bf p}+{\bf q}/2,-i\omega_m;{\bf r})
\nonumber
\\
&=&
\sum_{\sib p}
{
1-f(\varepsilon_{{\sib p}+{\sib q}/2}-\mu({\bf r}))
 -f(\varepsilon_{{\sib p}-{\sib q}/2}-\mu({\bf r}))
\over
\varepsilon_{{\sib p}+{\sib q}/2}+
\varepsilon_{{\sib p}-{\sib q}/2}-2\mu({\bf r})-i\nu_n
},
\label{eq.2.4}
\end{eqnarray}
where $f(\varepsilon)$ is the Fermi distribution function. Substituting eq. (\ref{eq.2.3b}) into eq. (\ref{eq.2.2}), we obtain
\begin{equation}
\Gamma({\bf q},i\nu_n;{\bf r})=-
{
U-g_{\rm r}^2D_0({\bf q},i\nu_n;{\bf r}) 
\over
1-[U-g_{\rm r}^2D_0({\bf q},i\nu_n;{\bf r})]\Pi({\bf q},i\nu_n;{\bf r})
}.
\label{eq.2.3c}
\end{equation}
We shall see that the effective attractive pairing interaction is given by 
\begin{equation}
U_{\rm eff}({\bf q},i\nu_n;{\bf r})=U-g_{\rm r}^2D_0({\bf q},i\nu_n;{\bf r}).
\label{eq.G1}
\end{equation}
\par
According to the Thouless criterion, $T_{\rm c}$ is given as the temperature at which the vertex function $\Gamma({\bf q},i\nu_n;{\bf r})$ in eq. (\ref{eq.2.3c}) develops a pole at ${\bf q}=\nu_n=0$. Within the LDA, since $\Gamma$ depends on position ${\bf r}$, this condition is satisfied at different temperatures $T_0({\bf r})$ depending on ${\bf r}$. In this approximation, we have to regard the highest $T_0({\bf r})$ as the phase transition temperature of the trapped gas.\cite{note} Since the density of particles is maximum at the center of the trap, we have $T_{\rm c}=T_0({\bf r}=0)$. Then the equation for $T_{\rm c}$ within the LDA is given by
\begin{equation}
1=U_{\rm eff}\Pi(0,0;{\bf r}=0),
\label{eq.2.5b}
\end{equation}
where
\begin{equation}
U_{\rm eff}\equiv U_{\rm eff}({\bf q}=0,i\nu_n=0;{\bf r}=0)
=U+{g_{\rm r}^2 \over 2\nu-2\mu}
\label{eq.G2}
\end{equation}
is the static part of the effective pairing interaction in the long wave length limit at ${\bf r}=0$. Equation (\ref{eq.2.5b}) indicates that even if the direct pairing interaction $U$ is weak, the effective pairing interaction $U_{\rm eff}$ in eq. (\ref{eq.G2}) can be large if $g_{\rm r}$ is large or when the chemical potential of the $b$-Boson ($\mu_B\equiv 2\mu$) approaches the bottom of the the Bose band $2\nu$. More explicitly, we note that eq. (\ref{eq.2.5b}) can be written in the well known BCS form
\begin{equation}
1=(U+g^2{1 \over 2\nu-2\mu})
\sum_{\sib p}
{\tanh(\varepsilon_{\sib p}-\mu)/2T_{\rm c} \over
2\varepsilon_{\sib p}-2\mu}.
\label{eq.2.5}
\end{equation}
\par
As shown in the literature,\cite{Baym,Kadanoff} the pole which develops at $T_{\rm c}$ at ${\bf q}=\nu_n=0$ signals an instability of the normal Fermi gas. At $T<T_{\rm c}$, the amplitude of this two-particle bound state can be shown to grow exponentially with time. This instability is removed by the appearance of a new phase below $T_{\rm c}$, described by the Cooper-pair order parameter.
\vskip3mm
\subsection{Fluctuation contributions to the chemical potential}
\vskip2mm
The chemical potential $\mu$ in eq. (\ref{eq.2.5}) is determined by the equation of the total number of Fermi atoms where a $b$-Boson also counts as two Fermi atoms. In the weak-coupling BCS theory in the absence of the $b$-Boson, this equation simply gives $\mu=\varepsilon_{\rm F}$ (where $\varepsilon_{\rm F}$ is the Fermi energy in the absence of any $b$-Bosons), assuming that the temperature dependence of $\mu$ can be neglected. Indeed, in most metallic (weak-coupling) superconductors, the deviation of $\mu(T_{\rm c})$ from $\varepsilon_{\rm F}$ is negligibly small as $\delta\mu\sim (T_{\rm c}/\varepsilon_{\rm F})^2\ll 1$.\cite{Mermin} On the other hand, when $g_{\rm r}$ is large or the Feshbach resonance strongly enhances the effective pairing interaction $U_{\rm eff}$, in analogy with the ``strong-coupling" superconductivity discussed by Nozi\`eres and Schmitt-Rink,\cite{Nozieres} the chemical potential $\mu$ is expected to deviate remarkably from $\varepsilon_{\rm F}$ due to the effect of fluctuations associated with the Cooper-pair channel. We now discuss these fluctuations.
\par
The equation giving the number of Fermi atoms determines the chemical potential $\mu$. This is obtained, as usual,\cite{Nozieres} from first calculating the thermodynamic potential $\Omega\equiv\int d{\bf r}\Omega({\bf r})$ and then using the identity $N=-\partial \Omega/\partial\mu$. The fluctuation contribution to the local thermodynamic potential $\Omega({\bf r})$ is shown in terms of diagrams in Fig. 2. In Fig. 2(a), we show the usual diagrams associated with the direct pairing interaction $-U$. Working to all orders in $U$, this gives\cite{Nozieres,Tokumitsu}
\begin{equation}
\delta\Omega_{U}({\bf r})={1 \over \beta}\sum_{{\sib q},\nu_n}e^{i\nu_n\delta}
\ln[1-U\Pi({\bf q},i\nu_n;{\bf r})].
\label{eq.G3}
\end{equation}
This is the contribution discussed in Ref.\cite{Nozieres}. In Fig. 2(b), we show the new fluctuation diagrams contributing to $\Omega({\bf r})$ associated with the Feshbach coupling interaction $g_{\rm r}$. Summing up to all orders, this gives
\begin{equation}
\delta\Omega_{g_{\rm r}}({\bf r})={1 \over \beta}\sum_{{\sib q},\nu_n}
e^{i\nu_n\delta}
\ln
\bigl[1+g_{\rm r}^2D_0({\bf q},i\nu_n;{\bf r}){\tilde \Pi}({\bf q},i\nu_n;{\bf r})],
\label{eq.G4}
\end{equation}
where 
\begin{equation}
{\tilde \Pi}({\bf q},i\nu_n;{\bf r})\equiv
{\Pi({\bf q},i\nu_n;{\bf r}) \over 1-U\Pi({\bf q},i\nu_n;{\bf r})}
\label{eq.G5}
\end{equation}
describes the particle-particle bubble including the effect of $U$ in the ladder approximation.
\par
We note that the total fluctuations contribution to $\Omega({\bf r})$ is the sum of the two contributions in eqs. ({\ref{eq.G3}) and (\ref{eq.G4}), which can be combined to give
\begin{eqnarray}
\delta\Omega({\bf r})
&=&
{1 \over \beta}\sum_{{\sib q},\nu_n}e^{i\nu_n\delta}
\ln
\Bigl[
\bigl[
1-U\Pi({\bf q},i\nu_n;{\bf r})
\bigr]
\bigr[
1+g_{\rm r}^2D_0({\bf q},i\nu_n;{\bf r})
{\tilde \Pi}({\bf q},i\nu_n;{\bf r})
\bigr]\Bigr]
\nonumber
\\
&=&
{1 \over \beta}\sum_{{\sib q},\nu_n}e^{i\nu_n\delta}
\ln[1-U_{\rm eff}({\bf q},i\nu_n;{\bf r})\Pi({\bf q},i\nu_n;{\bf r})],
\label{eq.G6}
\end{eqnarray}
where the effective pairing interaction $U_{\rm eff}({\bf q},i\nu_n;{\bf r})$ is as defined in eq. (\ref{eq.G1}). In other words, the only effect of the diagrams in Fig. 2(b) is to renormalize the pairing interaction as $U\to U_{\rm eff}({\bf q},i\nu_n;{\bf r})$ in eq. (\ref{eq.G3}). We also note that the collective fluctuations in eq. (\ref{eq.G6}) (which are Bosons) are precisely those which are associated with the poles of the particle-particle vertex function $\Gamma$ in eq. (\ref{eq.2.3c}). The thermodynamic potential for the contribution of the free Fermions and $b$-Bosons is given by the sum of\cite{AGD}
\begin{eqnarray}
\left\{
\begin{array}{l}
\displaystyle
\Omega^0_{F}({\bf r})=-{2 \over \beta}\sum_{{\sib p},\omega_m}
e^{i\omega_m\delta}
\ln G^{-1}_0({\bf p},i\omega_m;{\bf r}),
\\
\displaystyle
\Omega^0_{B}({\bf r})={1 \over \beta}\sum_{{\sib p},\nu_n}
e^{i\nu_n\delta}
\ln D^{-1}_0({\bf p},i\nu_n;{\bf r}).
\end{array}
\right.
\label{eq.G7}
\end{eqnarray}
\par
Combining all these results and using $N=-\partial\Omega/\partial\mu$, we find the result
\begin{equation}
N
=
\int d{\bf r}
\Bigl[
N_{\rm F}^0({\bf r})+2N_{\rm B}^0({\bf r})
-{1 \over \beta}\sum_{{\sib q},i\nu_n}
e^{i\delta\nu_n}
{\partial \over \partial\mu}
\rm{ln}
\bigl[
1-U_{\rm eff}({\bf q},i\nu_n;{\bf r})\Pi({\bf q},i\nu_n;{\bf r})
\bigr]
\Bigr],
\label{eq.2.8}
\end{equation}
where
\begin{eqnarray}
\left\{
\begin{array}{l}
\displaystyle
N^0_F={2 \over \beta}\sum_{{\sib p},\omega_m}
e^{i\omega_m\delta}
G_0({\bf q},i\omega_m;{\bf r})
=
2\sum_{\sib p}f(\varepsilon_{\sib p}-\mu({\bf r})),
\\
\displaystyle
N^0_B=-{1 \over \beta}\sum_{{\sib p},\nu_n}
e^{i\nu_n\delta}
D_0({\bf q},i\nu_n;{\bf r})
=
\sum_{\sib p}n_B(E_{\sib q}^0+2\nu-2\mu({\bf r})).
\end{array}
\right.
\label{eq.G8b}
\end{eqnarray}
Here $n_B(E)$ represents the Bose distribution function. Equations (\ref{eq.2.5b}) and (\ref{eq.2.8}) provide us with a closed set of two coupled equations, from which we can obtain $\mu(T,N)$ and $T_{\rm c}(N)$. In solving the coupled equations numerically, we have to take into account carefully the convergence factor $e^{i\nu_n\delta}$ in the last term in eq. (\ref{eq.2.8}). In the appendix, we explain how to sum up the Matsubara frequencies in this term. Above $T_{\rm c}$, we need only solve eq. (\ref{eq.2.8}) for $\mu(T,N)$. Our numerical results will be discussed in Sections IV and V. In the next section, however, we first discuss the physics involved in treating the fluctuations within the approximation developed in this section. This will allow us to better understand the nature of the Bosons involved in the crossover to a BEC regime.
\par
\vskip3mm
\section{Physics of Coupled $b$-Bosons and Cooper Pairs}
\par
The contribution arising from the interaction in eq. (\ref{eq.2.8}) can be described in terms of Bose fluctuations given by the solutions of
\begin{equation}
1=U_{\rm eff}({\bf q},\omega+i\delta;{\bf r})\Pi({\bf q},\omega+i\delta;{\bf r}),
\label{eq.G9}
\end{equation}
where we made the usual analytic continuation from the imaginary Bose Matsubara frequency to the real frequency axis as $i\nu_n\to\omega+i\delta$. While eq. (\ref{eq.G9}) appears relatively simple, it describes a complex situation involving the formation of preformed Cooper-pairs and their coupling to the bare $b$-Bosons in eq. (\ref{eq.2.1}) due to the Feshbach coupling interaction $g_{\rm r}$. The preformed Cooper-pairs and the $b$-molecules are coupled to each other, which leads to hybridization and only one kind of composite Bose molecule. It is useful to try to disentangle these effect, although there seems no unique way of doing this in such a coupled excitation problem.
\par
Denoting the number density of atoms given by the last term in eq. (\ref{eq.2.8}) as $\delta N_{\rm FL}({\bf r})$, we have
\begin{equation}
\delta N_{\rm FL}({\bf r})
={1 \over \beta}
\sum_{{\sib q},\nu_n}
e^{i\nu_n\delta}
\Bigl[
2g_{\rm r}^2\Pi D_0^2+U_{\rm eff}(q)
{\partial\Pi \over \partial\mu}
\Bigr]
\Bigl(
{1 \over
1-U_{\rm eff}(q)\Pi
}
\Bigr),
\label{eq.G10}
\end{equation}
where for simplicity we denote $\Pi({\bf q},i\nu_n;{\bf r})\equiv\Pi$, $D_0({\bf q},i\nu_n;{\bf r})\equiv D_0$, $U_{\rm eff}({\bf q},i\nu_n;{\bf r})\equiv U_{\rm eff}(q)$, and we have used the identity $\partial D_0/\partial\mu=-2D_0^2$. The effect of the particle-particle fluctuations described by $\Pi({\bf q},i\nu_n;{\bf r})$ naturally gives rise to a renormalized $b$-Boson propagator ${\tilde D}$ with a self-energy given by
\begin{equation}
\Sigma({\bf q},i\nu_n;{\bf r})=-g_{\rm r}^2{\tilde \Pi}({\bf q},i\nu_n;{\bf r}), 
\label{eq.G11b}
\end{equation}
where ${\tilde \Pi}$ has been defined in eq. (\ref{eq.G5}). Thus we have
\begin{eqnarray}
{\tilde D}^{-1}({\bf q},i\nu_n;{\bf r})
&\equiv&
D_0^{-1}+g_{\rm r}^2{\tilde \Pi}
\nonumber
\\
&=&
i\nu_n-[E_{\sib q}^0+2\nu-g_{\rm r}^2{\tilde \Pi}({\bf q},i\nu_n;{\bf r})]+2\mu({\bf r}).
\label{eq.G.11}
\end{eqnarray}
The self-energy $\Sigma$ in eq. (\ref{eq.G11b}) describes the effect of the decay of a $b$-Boson into two Fermi atoms and their recombination to a $b$-Boson, treating the non-resonant interaction $-U$ within the ladder approximation. This self-energy can be also obtained by summing up the diagrams shown in the third line in Fig. 1 (shaded bubble in the figure).
\par
One may verify that the first term in the square brackets in eq. (\ref{eq.G10}) renormalizes the number of $b$-Bosons, giving
\begin{eqnarray}
2N_{\rm B}^0({\bf r})
+{1 \over \beta}
\sum_{{\sib q},\nu_n}
e^{i\nu_n\delta}
{2g_{\rm r}^2\Pi D_0^2 \over
1-U_{\rm eff}(q)\Pi}
&=&
-{2 \over \beta}\sum_{{\sib q},\nu_n}
e^{i\nu_n\delta}
{(1-U\Pi)D_0 \over 1-U_{\rm eff}(q)\Pi}
\nonumber
\\
&=&
-{2 \over \beta}\sum_{{\sib q},\nu_n}
e^{i\nu_n\delta}
{\tilde D}({\bf q},i\nu_n;{\bf r})
\nonumber
\\
&\equiv&
2{\tilde N}_{\rm B}({\bf r}).
\label{eq.G.12}
\end{eqnarray}
\par
We recall that in our interacting Boson-Fermion model Hamiltonian, we have [see eq. (\ref{eq.AGY})]
\begin{equation}
N=N_{\rm F}+2N_{\rm B},
\end{equation}
where, within the LDA,
\begin{eqnarray}
\left\{
\begin{array}{l}
\displaystyle
N_{\rm F}=\int d{\bf r}\sum_{{\sib p},\sigma}
\langle c_{{\sib p}\sigma}^\dagger({\bf r}) c_{{\sib p}\sigma}({\bf r})
\rangle,
\\
\displaystyle
N_{\rm B}=\int d{\bf r}\sum_{\sib q}
\langle b_{\sib q}^\dagger({\bf r}) b_{\sib q}({\bf r})
\rangle
\end{array}
\right.
\label{eq.AG1}
\end{eqnarray}
give the number of Fermions and Bosons, respectively. As discussed in many body texts, $N_{\rm B}$ is defined in terms of the renormalized single particle $b$-Boson Green function ${\tilde D}({\bf q},i\nu_n;{\bf r})$. In fact, one sees that $N_{\rm B}={\tilde N}_{\rm B}\equiv\int d{\bf r}{\tilde N}_{\rm B}({\bf r})$, where ${\tilde N}_{\rm B}({\bf r})$ is defined in eq. (\ref{eq.G.12}). This means that $N_{\rm F}$ in eq. (\ref{eq.AG1}) is given by 
\begin{equation}
N_{\rm F}=N_{\rm F}^0+2N_{\rm C},
\label{eq.AG2}
\end{equation}
where $2N_{\rm C}$ is obtained from the second term in the square brackets in eq. (\ref{eq.G10}) as
\begin{equation}
2N_{\rm C}={1 \over \beta}\sum_{{\sib q},\nu_n}\int d{\bf r}
e^{i\nu_n\delta}
{\partial \Pi \over \partial\mu}
\Bigl(
{U_{\rm eff}(q) \over 1-U_{\rm eff}(q)\Pi}
\Bigr).
\label{eq.AG3}
\end{equation}
Equation (\ref{eq.AG2}) shows that the expression for $2N_{\rm C}$ in eq. (\ref{eq.AG3}) is the natural definition of the number of Cooper-pair Bosons which arise in the interacting Fermion gas with an attractive interaction given by $U_{\rm eff}(q)$.
\par
The preceding analysis leads to the following expression for $N$ given in Ref.\cite{Ohashi}:
\begin{equation}
N=N_{\rm F}^0+2N_{\rm B}+2N_{\rm C},
\label{eq.AG4}
\end{equation}
where $N_{\rm B}$ is the contribution from the $b$-Bosons and $N_{\rm C}$ is the contribution from the Cooper-pairs, both being strongly renormalized through the effect of the Feshbach resonance coupling $g_{\rm r}$. The problem with the decomposition given in eq. (\ref{eq.AG4}) is that when $g_{\rm r}\ne 0$, one must remember that the $b$-Bosons and the preformed Cooper-pairs are strongly hybridized. As a result, they give rise to a single Boson branch described by the pole of ${\tilde D}$ in eq. (\ref{eq.G.11}), namely
\begin{equation}
\omega_{\sib q}=
E_{\sib q}^0+2\nu-g_{\rm r}^2{\tilde \Pi}({\bf q},i\nu_n\to\omega_{\sib q}+i\delta;{\bf r})-2\mu({\bf r}).
\label{eq.G14}
\end{equation} 
One can understand the basic physics by going back to eq. (\ref{eq.G6}) and noting that
\begin{eqnarray}
1-U_{\rm eff}(q)\Pi
&=&
[1-U\Pi][1+g_{\rm r}^2D_0{\tilde \Pi}]
\nonumber
\\
&=&
[1-U\Pi]+g_r^2\Pi
{1 \over \omega-(E_{\sib q}^0+2\nu-2\mu({\bf r}))}.
\label{eq.G15}
\end{eqnarray}
This form shows clearly that the zeros of $1-U_{\rm eff}(q)\Pi$ correspond to a hybridized spectrum associated with the preformed Cooper-pair [$1-U\Pi=0$] and the $b$-Bosons [$\omega-(E_{\bf q}^0+2\nu-2\mu({\bf r}))=0$] by the coupling $g_{\rm r}^2\Pi$. The zeros of eq. (\ref{eq.G15}) are seen to be given by 
\begin{equation}
[1-U\Pi]\Bigl[\omega-[E_{\bf q}^0+2\nu-2\mu({\bf r})]\Bigr]+g_{\rm r}^2\Pi=0.
\label{eq.G16}
\end{equation}
One sees that the same hybridization physics is given by the poles of ${\tilde D}$, i.e.,
\begin{equation}
[\omega-(E_{\bf q}^0+2\nu-2\mu({\bf r}))]+
g_{\rm r}^2{\Pi \over 1-U\Pi}=0.
\label{eq.G17}
\end{equation}
All these versions emphasize that the renormalized Bosons given by ${\tilde D}^{-1}=0$ are the hybridized version of the bare preformed Cooper-pairs and the bare $b$-Bosons. One can still ask what is the relative ``weighting" of these two components in the renormalized Bose spectrum given by the pole of ${\tilde D}$, as given by the values of $N_{\rm B}$ and $N_{\rm C}$ in eq. (\ref{eq.AG4}). In Ref. \cite{Ohashi}, we calculated both these contributions as a function of the threshold $2\nu$ in the case of a uniform Fermi gas.
\par
Finally, we note that one can write
\begin{eqnarray}
1-U_{\rm eff}(q)\Pi
&=&
D_0\Bigl[D_0^{-1}-U_{\rm eff}(q)D_0^{-1}\Pi\Bigr]
\nonumber
\\
&=&
D_0\Bigl[D_0^{-1}-[{U \over D_0}-g_{\rm r}^2]\Pi\Bigr]
\nonumber
\\
&\equiv&
D_0D_{\rm M}^{-1}.
\label{eq.G18}
\end{eqnarray}
Putting this into eq. (\ref{eq.2.8}), one finds that the term $2N_{\rm B}^0$ is cancelled out by the $D_0$ factor in eq. (\ref{eq.G18}) and we are left with
\begin{equation}
N=\int d{\bf r}
[N_{\rm F}^0({\bf r})+2N_{\rm M}({\bf r})].
\label{eq.G19}
\end{equation}
Here the number density of composite molecular Bosons $N_{\rm M}({\bf r})$ is defined as
\begin{equation}
2N_{\rm M}({\bf r})\equiv-{1 \over \beta}
\sum_{{\sib q},\nu_n}
e^{i\nu_n\delta}
{\partial \over \partial \mu}
\ln D_{\rm M}^{-1},
\label{eq.G20}
\end{equation}
where $D_{\rm M}$ is defined in eq. (\ref{eq.G18}),
\begin{equation}
D_{\rm M}^{-1}({\bf q},i\nu_n;{\bf r})\equiv
i\nu_n-
[E_{\sib q}^0+2\nu+(U/D_0-g^2_{\rm r})\Pi]+2\mu({\bf r}).
\label{eq.G21}
\end{equation}
Note that $\Pi$ enters here, not ${\tilde \Pi}$ as in eq. (\ref{eq.G.11}). This composite molecule is a true hybrid of preformed Cooper-pairs and $b$-Bosons and cannot be associated with either component. However, one can verify that the poles of $D_{\rm M}$ are identical to those of ${\tilde D}$ as defined in eq. (\ref{eq.G.11}), in that [see eqs. (\ref{eq.G15}) and (\ref{eq.G18})]
\begin{eqnarray}
(1-U\Pi){\tilde D}^{-1}=D_0^{-1}-({U \over D_0}-g_{\rm r}^2)\Pi=D_{\rm M}^{-1}.
\label{eq.G22}
\end{eqnarray}
\par
\vskip3mm
\section{Transition Temperature in the BCS-BEC crossover with a Feshbach resonance}
\vskip2mm
\subsection{Renormalization for $U$ and $g_{\rm r}$}
\vskip2mm
In this section, we consider the case of an isotopic harmonic trap ($\omega_{0x}=\omega_{0y}=\omega_{0z}\equiv\omega_0$) for simplicity. It is convenient to take $\varepsilon_{\rm F}$ as the unit of energy (for one Fermi species). As for the unit of length, we take the radius $R_{\rm F}$ of a non-interacting Fermi gas at $T=0$ defined by $\varepsilon_{\rm F}(R_{\rm F})\equiv\varepsilon_{\rm F}-m\omega_0^2R_{\rm F}^2/2=0$, or $R_{\rm F}^2=2\varepsilon_{\rm F}/m\omega_0^2$. We also take, as the unit of the number of particles, the total number of Fermi atoms given by
\begin{equation}
N={4 \over \pi}\int_0^{R_{\rm F}}r^2dr
\int_0^{p_{\rm F}(r)}p^2dp
={(m\omega_0)^3R_{\rm F}^6 \over 24}, 
\label{eq.AG10}
\end{equation}
where $p_{\rm F}(r)=\sqrt{2m\varepsilon_{\rm F}(r)}$ is the Fermi momentum at $r$ within the LDA. With this normalization, since $N$ and $R_{\rm F}$ always appear with the coupling constants $U$ and $g_{\rm r}$ as ${\tilde U}\equiv UN/R_{\rm F}^3$ and ${\tilde g}_{\rm r}^2\equiv g_{\rm r}^2N/R_{\rm F}^3$, they can be absorbed into these coupling constants. (In the following, we simply write ${\tilde U}$ and ${\tilde g}_{\rm r}$ as $U$ and $g_{\rm r}$.) In our numerical calculations, we simply introduce a Gaussian cutoff $e^{-\varepsilon_{\sib p}^2/\omega_{\rm c}^2}$, with $\omega_{\rm c}=2\varepsilon_{\rm F}$, in the momentum summation involved in the particle-particle response function $\Pi({\bf q},i\nu_n;{\bf r})$ defined in eq. (\ref{eq.2.4}) and thus also in eq. (\ref{eq.2.5}).
\par
The interactions $-U$ and $g_{\rm r}$ used in this paper should be regarded as renormalized quantities effectively involving the effects of screening by (pseudo-spin and density) fluctuations and renormalization in terms of high energy processes: Gor'kov et al. showed within the mean-field BCS theory that in the dilute limit, $T_{\rm c}$ is suppressed by an effective interaction mediated by fluctuations described by particle-hole (p-h) bubble diagrams.\cite{Gorkov} This effect was also studied recently by Combescot\cite{Combescot} in connection with the possibility of high $T_{\rm c}$ superfluidity in $^6$Li. This reduction of $T_{\rm c}$ can be physically understood as a screening effect by (pseudo-) spin and density fluctuations, which weakens the pairing interaction between Fermi atoms. One should understand that $-U$ and $g_{\rm r}$ in our model in eq. (\ref{eq.2.1}) include this p-h screening.
\par
As for the renormalization of $-U$ and $g_{\rm r}$ by high energy processes, we note that our Fermi atomic gas has no physical cutoff-energy, in contrast to superconductivity in metals, where the lattice phonon Debye frequency acts the effective cutoff in the BCS gap equation. However, a prescription about the renormalization of the high energy interactions is implicit when we introduce a cutoff $\omega_c$ in the momentum summations. To see this, it is convenient to introduce an energy cutoff $E_c$. This cutoff energy is assumed to be much higher than $\omega_c$ and $T$, and it may be arbitrarily large. We can then divide the summation in the equation for $T_{\rm c}$ in eq. (\ref{eq.2.5}) into a low energy part $\varepsilon_{\sib p}=[0,\omega_c]$ and high energy part $\varepsilon_{\sib p}=[\omega_c,E_c]$. In this decomposition, when $E_c\gg\omega_c\gg|\mu|>T_{\rm c}$, we can take $\tanh(\varepsilon_{\sib p}-\mu)/2T_{\rm c}\to 1$ in the high energy part. Since this replacement corresponds to ignoring the influence of the Fermi distribution [recall that in eq. (\ref{eq.2.5}), $\tanh x/2T=1-2f(x)$], the high energy region can be described as two atoms interacting in a vacuum. The effect of the surrounding gas and Fermi statistics is absent. Then the decomposition of eq. (\ref{eq.2.5}) is reduced to\cite{noteY}
\begin{equation}
1=
U_{\rm eff}\sum_{[0,\omega_c]}
{\tanh(\varepsilon_{\sib p}-\mu)/2T_{\rm c} \over 2\varepsilon_{\sib p}-2\mu}+
U_{\rm eff}\sum_{[\omega_c,E_c]}{1 \over 2\varepsilon_{\sib p}-2\mu},
\label{eq.Y21}
\end{equation}
where $U_{\rm eff}=U+g_{\rm r}^2/(2\nu-2\mu)$. This equation can be rewritten as the form
\begin{eqnarray}
1
&=&
{U_{\rm eff} 
\over 1-U_{\rm eff}\sum_{[\omega_c,E_c]}{1 \over 2\varepsilon_{\sib p}-2\mu}}
\sum_{[0,\omega_c]}
{\tanh(\varepsilon_{\sib p}-\mu)/2T_{\rm c} \over
2\varepsilon_{\sib p}-2\mu}
\nonumber
\\
&\equiv&
U_{\rm eff}^{R}(\omega_c)
\sum_{[0,\omega_c]}
{\tanh(\varepsilon_{\sib p}-\mu)/2T_{\rm c} \over
2\varepsilon_{\sib p}-2\mu}.
\label{eq.Y22}
\end{eqnarray}
In this equation for $T_{\rm c}$, high energy two-particle scattering processes in the region $[\omega_c,E_c]$ has been absorbed into the renormalized coupling $U_{\rm eff}^R(\omega_c)=U^R(\omega_c)+{g_{\rm r}^R}^2(\omega_c)/(2\nu^R(\omega_c)-2\mu)$ describing the low energy physics in the region $[0,\omega_c]$. The renormalized interaction $U^R(\omega_c)$, Feshbach coupling $g_{\rm r}^R(\omega_c)$ and threshold energy $2\nu^R(\omega_c)$ are thus given by
\begin{eqnarray}
\left\{
\begin{array}{l}
\displaystyle
U^R(\omega_c)=
{U \over 1-U\sum_{[\omega_c,E_c]}{1 \over 2\varepsilon_{\sib p}-2\mu}},
\\
\displaystyle
g_{\rm r}^R(\omega_c)=
{g_{\rm r} \over 1-U\sum_{[\omega_c,E_c]}{1 \over 2\varepsilon_{\sib p}-2\mu}},
\\
\displaystyle
2\nu^R(\omega_c)=2\nu-g_r^2
{\sum_{[\omega_c,E_c]}{1 \over 2\varepsilon_{\sib p}-2\mu} \over 
1-U\sum_{[\omega_c,E_c]}{1 \over 2\varepsilon_{\sib p}-2\mu}}.
\end{array}
\right.
\label{eq.Y23}
\end{eqnarray}
Physically, $U^R(\omega_c)$ is the non-resonant Fermion-Fermion interaction enhanced by high energy scattering processes and is diagrammatically described by Fig. 3 (a). The renormalized Feshbach coupling $g_{\rm r}^R(\omega_c)$ involves the three point vertex correction coming from the high-energy processes in Fig. 3(b). Finally, the renormalized threshold energy $2\nu^R(\omega_c)$ originates from the self-energy correction caused by the break of a Bose molecule into two Fermi atoms with high momenta, which can be described by the diagrams in Fig. 3(c).
\par
We should determine the renormalized quantities experimentally. In this regard, the renormalized quantities for the low energy limit ($\omega_c=0$) can be introduced.\cite{Kokkelman,Milstein} One can write the renormalized parameters in eq. (\ref{eq.Y23}) in terms of their low energy values ($\omega_c=0$) as follows:
\begin{eqnarray}
\left\{
\begin{array}{l}
\displaystyle
U^R(\omega_c)=
{U^R(0) \over 1+U^R(0)\sum_{[0,\omega_c]}{1 \over 2\varepsilon_{\sib p}-2\mu}},
\\
\displaystyle
g_{\rm r}^R(\omega_c)=
{g_{\rm r}^R(0) \over 1+U^R(0)\sum_{[0,\omega_c]}{1 \over 2\varepsilon_{\sib p}-2\mu}},
\\
\displaystyle
2\nu^R(\omega_c)=2\nu^R(0)+g_r^R(0)^2
{\sum_{[0,\omega_c]}{1 \over 2\varepsilon_{\sib p}-2\mu} \over 
1+U^R(0)\sum_{[0,\omega_c]}{1 \over 2\varepsilon_{\sib p}-2\mu}}.
\end{array}
\right.
\label{eq.Y23b}
\end{eqnarray}
Written in terms of the renormalized variables given in eq. (\ref{eq.Y23b}), the BCS gap equation in eq. (\ref{eq.Y22}) no longer involves $E_c$ explicitly. Although we simply write the couplings as $-U$, $g_{\rm r}$ and the threshold energy as $2\nu$ in our model, we can regard them as renormalized quantities incorporating the high energy two-particle scattering processes, as discussed above.
\par
The present way of renormalization can reproduce the cutoff-free theories in Refs.\cite{Melo,Randeria,Pethick2}. When we write eq. (\ref{eq.Y22}) using the renormalized variables in terms of the low energy value $U_{\rm eff}^R(\omega_{\rm c}=0)$ as given by eq. (\ref{eq.Y23b}), we obtain
\begin{equation}
1=
{U^R_{\rm eff}(0) 
\over 1+U^R_{\rm eff}(0)\sum_{[0,\omega_c]}{1 \over 2\varepsilon_{\sib p}-2\mu}}
\sum_{[0,\omega_c]}
{\tanh(\varepsilon_{\sib p}-\mu)/2T_{\rm c} \over
2\varepsilon_{\sib p}-2\mu}.
\label{eq.Y24}
\end{equation}
This equation can be in turn rewritten in the form
\begin{eqnarray}
1&=&
{U^R_{\rm eff}(0) 
\sum_{[0,\omega_c]}
\Bigl[
{\tanh(\varepsilon_{\sib p}-\mu)/2T_{\rm c} \over
2\varepsilon_{\sib p}-2\mu}
-
{1 \over 2\varepsilon_{\sib p}-2\mu}}
\Bigr]
\nonumber
\\
&\simeq&
{U^R_{\rm eff}(0) 
\sum_{[0,\infty]}
\Bigl[
{\tanh(\varepsilon_{\sib p}-\mu)/2T_{\rm c} \over
2\varepsilon_{\sib p}-2\mu}
-
{1 \over 2\varepsilon_{\sib p}-2\mu}}
\Bigr].
\label{eq.Y25}
\end{eqnarray}
Since the two terms in the square brackets almost cancel out each other in the high energy region $[\omega_c,\infty]$ (we assume that $\omega_c\gg T$), we can eliminate the finite upper limit $\omega_c$ in the summation. This cutoff-free expression is similar to the one used in Ref.\cite{Pethick2}. The cutoff-free BCS equation for $T_{\rm c}$ used by Randeria et al.\cite{Melo,Randeria} can be also obtained from eq. (\ref{eq.2.5}). Writing it as
\begin{equation}
1=U_{\rm eff}\sum_{[0,E_c]}\Bigl[
{\tanh (\varepsilon_{\sib p}-\mu)/T_{\rm c} 
\over 
2\varepsilon_{\sib p}-\mu}
-
{1 \over 2\varepsilon_{\sib p}}
+
{1 \over 2\varepsilon_{\sib p}}
\Bigr],
\label{eq.Y26}
\end{equation}
we can turn this into
\begin{equation}
1={\bar U}_{\rm eff}\sum_{[0,E_c]}\Bigl[
{\tanh (\varepsilon_{\sib p}-\mu)/T_{\rm c} 
\over 
2\varepsilon_{\sib p}-\mu}
-
{1 \over 2\varepsilon_{\sib p}}
\Bigr],
\label{eq.Y27}
\end{equation}
where the renormalized interaction ${\bar U}_{\rm eff}={\bar U}+{\bar g}_{\rm r}^2/(2{\bar \nu}-2\mu)$ is defined by
\begin{eqnarray}
\left\{
\begin{array}{l}
\displaystyle
{\bar U}={U \over 1-U\sum_{[0,E_c]}{1 \over 2\varepsilon_{\sib p}}},
\\
\displaystyle
{\bar g}_{\rm r}=g_{\rm r}
{1 \over 1-U\sum_{[0,E_c]}{1 \over 2\varepsilon_{\sib p}}},
\\
\displaystyle
2{\bar \nu}=2\nu-g_{\rm r}^2
\sum_{\sib p}
{\sum_{[0,E_c]}{1 \over 2\varepsilon_{\sib p}}
\over 1-U\sum_{[0,E_c]}{1 \over 2\varepsilon_{\sib p}}}.
\end{array}
\right.
\label{eq.Y28}
\end{eqnarray}
All these expressions for the equation for $T_{\rm c}$ give physically the same result, assuming that one takes the appropriate values for renormalized parameters.
\par
With regard to the renormalization of the coupling constants, recent work by Kokkelmans et al.\cite{Kokkelman} has discussed the coupled Fermion-Boson model given in eq. (\ref{eq.2.1}) and incorporated the full two-body scattering theory into  an improved fashion. Milstein et al.\cite{Milstein} have more recently derived identical equations to eqs. (4) and (6) in Ref. \cite{Ohashi} (these correspond to eqs. (\ref{eq.2.5}) and (\ref{eq.2.8}) for a trapped Fermi gas), which include the effect of particle-particle fluctuations and the appearance of Cooper-pairs. They have extended our analysis in Ref. \cite{Ohashi} by renormalizing these equations in a way which treats the two-body scattering and bound states correctly. Their work thus leads to explicit numerical values for the various parameters in the renormalized model corresponding to eq. (\ref{eq.2.1}). 
\par
In the rest of this Section and Sections V and VI, we discuss our results for $T_{\rm c}$ and the region above, using $U=0.3\varepsilon_{\rm F}$ and $g_{\rm r}=0.6\varepsilon_{\rm F}$. Our diagrammatic approximation for the fluctuations, as summarized in Figs. 1 and 2, assumes that the interactions are weak. In Section VII, we discuss the analogous results predicted by our equations (\ref{eq.2.5}) and (\ref{eq.2.8}) using a value $g_{\rm r}\gg\varepsilon_{\rm F}$, such as found in Refs. \cite{Holland,Milstein}. In this case, the Feshbach resonance is very broad and thus can have a huge effect on $T_{\rm c}$ even if the threshold $2\nu\gg 2\varepsilon_{\rm F}$. 
\par
\vskip2mm
\subsection{Crossover behavior at $T_{\rm c}$ within LDA}
\vskip2mm
Our numerical solution for $T_{\rm c}$ and $\mu(T_{\rm c},N)$ from eqs. (\ref{eq.2.5}) and (\ref{eq.2.8}) for a trapped Fermi gas gives results which are very similar to those for a uniform Fermi gas obtained in Ref. \cite{Ohashi}. Figure 4 shows the BCS-BEC crossover in a trapped Fermi gas. As the threshold energy $2\nu$ is lowered, Fig. 4(a) shows that $T_{\rm c}$ deviates from the mean-field BCS theory (`BCS' in the figure) and approaches the BEC phase transition temperature of a gas of $N/2$ free Bosons of mass $M=2m$ (`BEC' in the figure). The chemical potential $\mu$ also changes remarkably in this crossover as shown in Fig. 4(b). As expected from the mean-field BCS theory, we obtain $\mu\simeq\varepsilon_{\rm F}$ for $2\nu\gesim2\varepsilon_{\rm F}$ (BCS regime) while $\mu$ deviates from $\varepsilon_{\rm F}$ and approaches $\nu$ for $\nu\lesssim 0$ (BEC regime). The latter result reflects that the $b$-Bosons become the dominant contribution in the BEC regime. As a result, the phase transition occurs when the chemical potential of the $b$-Boson $\mu_{\rm B}$ ($=2\mu$) reaches the bottom of the Bose band ($2\nu$) at the center of the trap (within our LDA). We note that we can rewrite the equation for $T_{\rm c}$ in eq. (\ref{eq.2.5b}) in the form
\begin{equation}
2\nu=2\mu+g_{\rm r}^2{\tilde \Pi}(0,0,{\bf r}=0).
\label{eq.3.1}
\end{equation}
This equation reduces to $2\nu=2\mu$ $(=\mu_{\rm B})$ in the BEC limit, where the Fermi atoms and hence the particle-particle fluctuations described by $\Pi$ are absent. Since, in this BEC regime, the dominant excitations are $N/2$ free Bosons, we find the usual result for a trapped free Bose gas $T_{\rm c}=\omega_0[N/2\zeta(3)]^{1/3}$,\cite{Pethick} where $\omega_0$ is the trap frequency and $\zeta(x)$ is the usual zeta function (we set $\hbar=1$). Using the fact that $T_{\rm F}=(6N_\uparrow)^{1/3}\omega_0$ is the Fermi temperature for a trapped free Fermi gas (with $N_\uparrow=N/2$), we can write this expression for $T_{\rm c}$ in terms of $T_{\rm F}$ to give
\begin{equation}
T_{\rm c}=
\Bigl(
{1 \over 6\zeta(3)}
\Bigr)^{1/3}\varepsilon_{\rm F}=0.518T_{\rm F}.
\label{eq.3.3}
\end{equation}
Figure 4(a) shows that this $T_{\rm c}$ is obtained in the BEC regime as the limiting maximum value in the trapped Fermi gas. We recall that Fig. 4 is for $U=0.3\varepsilon_{\rm F}$ and $g_{\rm r}=0.6\varepsilon_{\rm F}$.
\par
We note that this upper limit for $T_{\rm c}$ is much higher than that of a uniform Fermi gas in the absence of a trap.\cite{Ohashi} In the uniform system, $T_{\rm c}$ in the BEC regime is given by\cite{Ohashi,Randeria} 
\begin{equation}
T_{\rm c}=2
\Bigl(
{1 \over 6\sqrt{\pi}\zeta(3/2)}
\Bigr)^{2/3}\varepsilon_{\rm F}=0.218T_{\rm F},
\label{eq.3.2}
\end{equation}
where the Fermi temperature in the uniform system is given by $T_{\rm F}=(6\pi^2 N_\uparrow)^{2/3}/2m$.\cite{Mermin} One reason for this different maximum $T_{\rm c}$'s between the two cases comes from the different definition of $T_{\rm F}$. In addition, one should note that the harmonic trap potential enhances $T_{\rm c}$ through the effective density of states (DOS) of Bose molecules. When we carry out the spatial integration in the equation for the total number of trapped particles in the BEC regime [$N=2\int d{\bf r}N_{\rm B}({\bf r})$], this equation is reduced to\cite{Pethick}
\begin{equation}
1=3\int_0^\infty dE E^2 {1 \over e^{{\bar \beta} E}-1}~~~~~(T=T_{\rm c}),
\label{eq.3.4}
\end{equation}
where ${\bar \beta}\equiv T_{\rm F}/T$ and the energy in the integral is normalized in units of $\varepsilon_{\rm F}$. The corresponding equation in the uniform system is given by
\begin{equation}
1=3\sqrt{2}\int_0^\infty dE \sqrt{E} {1 \over e^{{\bar \beta} E}-1}.\label{eq.3.5}
\end{equation}
This shows that the low energy region of the DOS has larger weight in the uniform gas (DOS $\propto \sqrt{E}$) than in the trapped one (DOS $\propto E^2$). As a result, the transition temperature (relative to $T_{\rm F}$ for each case) must be higher in the trapped gas than in the uniform gas.
\vskip3mm
\section{Long-lived Composite Molecular Bosons}
\vskip3mm
\subsection{Damping of $b$-Boson and condition for stable molecules at $T_{\rm c}$}
\vskip2mm
The decay of $b$-Boson into two Fermi atoms associated with the Feshbach resonance is described by the imaginary part of the (analytic-continued) self-energy of the renormalized Bose Green function in eq. (\ref{eq.G.11}) 
\begin{equation}
\gamma\equiv -Im\Sigma({\bf q},i\nu_n\to\omega+i\delta;{\bf r})=g_{\rm r}^2Im{\tilde \Pi}({\bf q},\omega+i\delta;{\bf r}). 
\label{eq.gamma}
\end{equation}
Since $Im\Pi({\bf q},\omega+i\delta;{\bf r})$ is given by
\begin{equation}
Im \Pi({\bf q},\omega+i\delta;{\bf r})=
{m\sqrt{m} \over 4\pi\beta\sqrt{E^0_{\sib q}}}
\Theta(\omega+2\mu({\bf r})-E^0_{\sib q})
\ln
{
\cosh{\beta \over 2}({\omega \over 2}
+\sqrt{(\omega+2\mu({\bf r})-E^0_{\sib q})E^0_{\sib q}})
\over
\cosh{\beta \over 2}({\omega \over 2}
-\sqrt{(\omega+2\mu({\bf r})-E^0_{\sib q})E^0_{\sib q}})
},
\label{eq.3.6}
\end{equation}
where $\Theta(x)$ is the step function, the damping rate $\gamma$ is finite only in the region $\omega\ge \omega_{\rm th}\equiv E^0_{\sib q}-2\mu({\bf r})$. This step function restriction is clear when we note that the denominator of eq. (\ref{eq.2.4}) can be written as $[i\nu_n+2\mu({\bf r})-E^0_{\sib q}]-2\varepsilon_{\sib p}$. At ${\bf q}=0$, eq. (\ref{eq.3.6}) can be simplified to
\begin{equation}
Im \Pi({\bf q}=0,\omega+i\delta;{\bf r})=
{m\sqrt{m} \over 4\pi}
\Theta(\omega+2\mu({\bf r}))
\sqrt{\omega+2\mu({\bf r})}
\tanh{\beta\omega \over 4}.
\label{eq.3.6b2}
\end{equation}
Since the energy to dissociate one $b$-Boson (with center of mass momentum ${\bf q}$) into two Fermi atoms is given by $[\varepsilon_{{\sib p}+{\sib q}/2}-\mu({\bf r})]+[\varepsilon_{-{\sib p}+{\sib q}/2}-\mu({\bf r})]$, the threshold energy $\omega_{\rm th}$ is found to correspond to the {\it minimum} energy to destroy one $b$-Boson with ${\bf q}=0$. 
\par
The damping rate $\gamma$ leads to finite lifetime of $b$-Boson, or equivalently gives a finite width to the peak in the spectral density of the (renormalized) Bose excitation spectrum given by
\begin{equation}
\rho_{\rm B}({\bf q},\omega;{\bf r})\equiv
-{1 \over \pi}Im{\tilde D}({\bf q},i\nu_n\to \omega+i\delta; {\bf r}).
\label{eq.Boson1}
\end{equation}
In the BCS regime, where $\gamma$ is always finite for $\omega>0$ (see the inset in Fig. 5(a)), the quasi-particle $b$-Bose spectral density shows a broad peak, as shown in Fig. 5(a). This $b$-Boson thus has a finite lifetime, which is given by the inverse of the peak width, due to decay into two Fermi atoms by the coupling to the Feshbach resonance. As one approaches the BEC regime, $\omega_{\rm th}>0$ is realized (as shown in the insets in Figs. 5(b) and 5(c)). In this case, when the $b$-Boson pole given by eq. (\ref{eq.G14}),
\begin{equation}
\omega_{\sib q}=\bigl[E_{\sib q}^0-2\mu({\bf r})\bigr]+\bigl[2\nu-g_{\rm r}^2{\tilde \Pi}({\bf q},\omega_{\sib q}+i\delta;{\bf r})\bigr], 
\label{eq.3.6b}
\end{equation}
appears below $\omega_{\rm th}$, its lifetime is infinite. It thus appears as a $\delta$-function peak in the $b$-Boson excitation spectral density $\rho_{\rm B}$ given by eq. (\ref{eq.Boson1}). We show this case in Figs. 5(b) and 5(c). In Fig. 5(b), the excitation spectrum is still accompanied by a incoherent part in high energy region ($\omega>\omega_{\rm th}$), in addition to the undamped bound state pole below $\omega_{\rm th}$. On the other hand, the incoherent part is seen to be almost absent in the BEC regime in Fig. 5(c), evidence that a stable $b$-molecule dominates the two-particle excitation spectrum.
\par
The condition for such a stable $b$-Boson is simply that\cite{Ohashi}
\begin{equation}
2{\tilde \nu}\equiv 2\nu-g_{\rm r}^2{\tilde \Pi}({\bf q},\omega_{\sib q}+i\delta;{\bf r})<0~~~~~[E_{\sib q}^0-2\mu({\bf r})>0].
\label{eq.3.7}
\end{equation}
This means that the renormalized threshold energy defined by $2{\tilde \nu}$ in eq. (\ref{eq.3.7}) must be negative for the stable undamped $b$-Bose molecules to appear.\cite{Ohashi} In the region $\omega={\bf q}=0$, this condition can be approximately written as $2{\tilde \nu}\simeq2\nu-g_{\rm r}^2{\tilde \Pi}(0,0;{\bf r})<0$. Since $\Pi(0,0,{\bf r})$ can be shown to be positive and also decreases as ${\bf r}$ increases, the condition $2{\tilde \nu}<0$ is first satisfied at the center of the trap.
\par
We can divide the number $N_{\rm B}$ of renormalized $b$-Bosons in eq. (\ref{eq.AG4}) into the contribution from stable long-lived part ($N_{\rm B}^{\gamma=0}$) and a Feshbach resonating contribution with a finite lifetime (denoted by $N_{\rm B}^{\gamma>0}$). In order to evaluate the contribution from the stable part $N_{\rm B}^{\gamma=0}$, it is convenient to introduce the spectral representation of the renormalized Bose Green function as\cite{Nozieres}
\begin{equation}
{\tilde D}({\bf q},i\nu_n;{\bf r})=
-{1 \over \pi}\int_{-\infty}^\infty dz
{1 \over i\nu_n-z}Im{\tilde D}({\bf q},i\nu_n\to z+i\delta;{\bf r}).
\label{eq.Y1}
\end{equation}
Using this expression, the summation in terms of the Boson Matsubara frequency in $N_{\rm B}$ in eq. (\ref{eq.AG4}) can be carried out to give
\begin{equation}
N_{\rm B}=
-{1 \over \pi}\sum_{\sib q}
\int d{\bf r}
\int_{-\infty}^\infty dz
n_B(z)Im{\tilde D}({\bf q},z+i\delta;{\bf r}).
\label{eq.Y2}
\end{equation}
Using this, the contribution from the stable poles (with no imaginary part) is given by 
\begin{eqnarray}
N_{\rm B}^{\gamma=0}
&=&
\int d{\bf r}\sum_{\sib q}^{\rm poles}
\int_{-\infty}^\infty dz
\delta (z-[E_{\sib q}-2\mu({\bf r})]
-[2\mu-g_{\rm r}^2{\tilde \Pi}
({\bf q},z+i\delta;{\bf r})])
\nonumber
\\
&=&
\int d{\bf r}
\sum_{\sib q}^{\rm poles}Z({\bf q};{\bf r})n_{\rm B}
(\omega_{\bf q}).
\label{eq.3.8}
\end{eqnarray}
Here, $Z({\bf q};{\bf r})$ arises from what is called mass renormalization and is given by
\begin{equation}
Z({\bf q};{\bf r})^{-1}\equiv 1+g_{\rm r}^2
{\partial{\tilde \Pi}({\bf q},\omega_{\bf q},{\bf r}) \over \partial \omega_{\sib q}}.
\label{eq.3.8b}
\end{equation} 
The sum in eq. (\ref{eq.3.8}) is over the solutions $\omega_{\sib q}$ satisfying eq. (\ref{eq.3.6b}).
\par
Besides this Feshbach component of the two-particle bound state given by eq. (\ref{eq.3.8}), there is a component from Cooper-pair associated with $N_{\rm C}$ as defined in eq. (\ref{eq.AG3}). One can rewrite eq. (\ref{eq.AG3}) in the form
\begin{equation}
2N_{\rm C}=
{1 \over \beta}
\sum_{{\sib q},\nu_n}
\int d{\bf r}
{{\tilde D} \over D_0}
{U_{\rm eff}(q) \over 1-U\Pi}
{\partial \Pi \over \partial \mu}.
\label{eq.3.9}
\end{equation}
We next use the fact that 
\begin{equation}
{\partial{\tilde \Pi} \over \partial\mu}
=
\Bigl(
{1 \over 1-U\Pi}
\Bigr)^2
{\partial\Pi \over \partial\mu},
\label{eq.3.9b}
\end{equation}
and that for $\omega=\omega_{\sib q}$ [given by eq. (\ref{eq.3.6b})], one has the identities $D_0^{-1}=-g_{\rm r}^2{\tilde \Pi}$ and $U_{\rm eff}(q)=\Pi^{-1}$. With these relations, the contribution $N_{\rm C}^{\gamma=0}$ of undamped Bosons associated with Cooper pairing can be reduced to the expression\cite{Ohashi}
\begin{equation}
N_{\rm C}^{\gamma=0}=
{g_{\rm r}^2 \over 2}\int d{\bf r}
\sum_{\sib q}^{\rm poles}
{\partial {\tilde \Pi}({\bf q},\omega_{\sib q};{\bf r}) \over \partial\mu}Z({\bf q};{\bf r})n_{\rm B}(\omega_{\sib q}),
\label{eq.3.11}
\end{equation}
where $Z({\bf q};{\bf r})$ has been defined in eq. (\ref{eq.3.8b}). There is also a contribution from scattering states and hence we have $N_{\rm C}=N_{\rm C}^{\gamma=0}+N_{\rm C}^{\rm sc}$, where $N_{\rm C}^{\rm sc}$ represents the scattering contribution associated with damped particle-particle (or Cooper channel) fluctuations. 
\par
Figure 4(c) shows the change of the character of the excitations in the BCS-BEC crossover in the trap. As $\nu$ is lowered and approaches $\varepsilon_{\rm F}$, the Feshbach resonance molecular state starts to play a role and the Feshbach damped molecules ($N_{\rm B}^{\gamma>0}$) appear. These Bosons are replaced by the stable molecules ($N_{\rm B}^{\gamma=0}$) around $\nu=0.18\varepsilon_{\rm F}$ for the parameters used in Fig. 4(c). At the same time, a contribution from preformed Cooper-pairs ($N_{\rm C}^{\gamma=0}$) appear. In the BEC regime, stable Feshbach molecules become dominant and $N_{\rm B}^{\gamma=0}$ approaches $0.5$ ($=N/2$). The contribution from preformed Cooper-pairs are seen to again disappear in the BEC limit $\nu<0$. The increase in the total number of molecules is, however, slower than that found in the uniform system.\cite{Ohashi} This is simply because the stable molecules initially only begin to appear at the center of the trap, rather than throughout the system as in the uniform gas case..
\par
Figure 6 shows the number of composite Bosons $N_{\rm M}\equiv\int d{\bf r}N_{\rm M}({\bf r})$ as a function of $\nu$ [where $N_{\rm M}({\bf r})$ is defined in eq. (\ref{eq.G20})]. Although each component of Boson ($N_{\rm B}^{\gamma=0}$, $N_{\rm B}^{\gamma>0}$, $N_{\rm C}^{\gamma=0}$ and $N_{\rm C}^{\rm sc}$) shows a singular behavior around $\nu=0.18\varepsilon_{\rm F}$ in Fig. 4(c), $N_{\rm M}$~($\equiv N_{\rm B}^{\gamma=0}+N_{\rm B}^{\gamma>0}+N_{\rm C}^{\gamma=0}+N_{\rm C}^{\rm sc}$) is found to increase smoothly as the threshold energy $2\nu$ is lowered, as expected. 
\par
As we have discussed in detail in Section III, the strong hybridization induced by the Feshbach coupling interaction $g_{\rm r}$ means that there is only a single two-particle Bose branch, whose energy is given by the solution of eq. (\ref{eq.3.6b}). Thus $N_{\rm B}^{\gamma=0}$ and $N_{\rm C}^{\gamma=0}$ only represent the relative spectral weights of the two components (corresponding to long-lived $b$-Bosons and stable preformed Cooper-pairs) of this composite Boson. This decomposition is only meant as a way of understanding the physics of the BCS-BEC crossover which arises in the model given by eq. (\ref{eq.2.1}). This interpretation is, however, very useful to understand the difference between the present BCS-BEC crossover and that in strong-coupling superconductivity with no Feshbach resonance: In the latter case, the preformed Cooper-pairs are the dominant excitations in the strong coupling BEC regime.\cite{Nozieres,Melo,Randeria} In the present case, for the values of the parameters $U=0.3\varepsilon_{\rm F}$ and $g_{\rm r}=0.6\varepsilon_{\rm F}$ used in Fig. 4, although preformed Cooper-pairs make a contribution, the stable Feshbach molecules dominate the excitation spectrum in the BEC limit ($\nu<0$). 

\par
\vskip3mm
\subsection{The binding energy of composite Bosons}
\vskip2mm
The binding energy of the long-lived composite Boson is the energy to break this molecule into two Fermi atoms. This dissociation energy can be evaluated from the energy difference between the bound state energy $\omega_{\bf q}$ in eq. (\ref{eq.3.6b}) and the excitation spectrum of the free Fermi gas measured from the chemical potential $\mu$. In particular, the binding energy of the zero-energy ($\omega_{{\sib q}=0}=0$) two-particle bound state with ${\bf q}=0$ which appears at $T_{\rm c}$ is given by
\begin{eqnarray}
E_{\rm binding}=2[\varepsilon_{{\sib q}=0}-\mu({\bf r}=0)]-\omega_{{\sib q}=0}
&=&
2|\mu|
\nonumber
\\
&=&
2|{\tilde \nu}({\bf q}=0,\omega_{\sib q}=0;{\bf r}=0)|,
\label{eq.Y10}
\end{eqnarray}
where we have used $\mu({\bf r}=0)<0$ and eq. (\ref{eq.3.7}) to obtain expressions on the right hand side. This result means that $E_{\rm binding}$ is equal to the absolute value of the energy of a renormalized $b$-molecule at the bottom of the renormalized Bose spectrum. 
\par
The threshold energy $\omega_{\rm th}({\bf q}=0)$ of the two-particle spectrum is defined as where $Im\Pi({\bf q}=0,\omega;{\bf r}=0)$ given in eq. (\ref{eq.3.6}) becomes finite (see Fig. 5). We recall that $Im\Pi$ describes the excitation spectrum of the Cooper-pair channel fluctuations. This frequency $\omega_{\rm th}({\bf q}=0)$ is the same as the threshold energy of the continuum spectrum of the renormalized $b$-Boson self-energy in eq. (\ref{eq.G11b}), since $\gamma=-Im \Sigma=g_{\rm r}^2Im{\tilde \Pi}(i\nu_n\to \omega+i\delta)\propto Im\Pi(\omega+i\delta)$. When a stable $b$-Boson at $\omega_{{\sib q}=0}=0$ is excited to a state with energy $E\ge \omega_{\rm th}$, this excited $b$-Boson will decay into two Fermi atoms lading to a finite lifetime for this excited state. We conclude that $\omega_{\rm th}({\bf q}=0)=E_{\rm binding}$, i.e., the particle-particle threshold energy $\omega_{\rm th}({\bf q}=0)$ is equal to the binding energy of a stable two-particle bound state.
\par
Since the chemical potential $\mu$ approaches $\nu$ in the BEC limit as shown in Fig. 4(b), we find from eq. (\ref{eq.Y10}) that $E_{\rm binding}\to 2|\nu|$ in this regime. This energy $2|\nu|$ is just equal to the energy needed to transfer a free $b$-Boson at the bottom of the Bose energy band ($E_{{\sib q}=0}^0+2\nu=-2|\nu|$) to two Fermi atoms with zero energy ($2\varepsilon_{{\sib q}=0}=0$). This result is consistent with the fact that the preformed Cooper-pair component in the bound state is absent in the BEC limit, leaving the $b$-molecule component.
\par
The binding energy of the preformed Cooper-pair in the strong-coupling BEC regime has been also discussed in superconductivity,\cite{Nozieres,Melo,Randeria,Leggett} and it is useful to understand how the present results are related to this previous work. To see this, let us recall the cutoff-free equation for $T_{\rm c}$ given by eq. (\ref{eq.Y27}). Since $\mu\ll 0$ in the BEC limit, one can approximately take $\tanh(\varepsilon_{\sib p}-\mu)/2T=1$. Summing over ${\bf p}$ in eq. (\ref{eq.Y27}), we obtain
\begin{equation}
1=-\Bigl({\bar U}+{\bar g}_{\rm r}^2{1 \over 2{\bar \nu}-2\mu}\Bigr)
{m\sqrt{2m|\mu|} \over 4\pi}.
\label{eq.AG41}
\end{equation}
When we write the renormalized attractive interaction as ${\bar U}_{\rm eff}\equiv {\bar U}+{\bar g}_{\rm r}^2/(2{\bar \nu}-2\mu)\equiv -4\pi a_s/m$, where $a_s$ is an $s$-wave scattering length, we find $|\mu|=1/2ma_s^2$, which is formally the same expression as that obtained in strong-coupling superconductivity discussions.\cite{Melo,Randeria,Leggett} In the present model, where the pairing interaction mediated by the Feshbach resonance is dominant in the BEC regime, we can neglect the non-resonant part ${\bar U}$ in ${\bar U}_{\rm eff}$. Then in the case of ${\bar \nu}<0$, the solution of eq. (\ref{eq.AG41}) is\cite{note33}}
\begin{equation}
{\mu \over {\bar \nu}}=1+{m^3{\bar g}_{\rm r}^4 \over 64\pi^2{\bar \nu}}
\Bigl[
\sqrt{1+{128\pi^2|{\bar \nu}| \over m^3{\bar g}_{\rm r}^4}}
-1
\Bigr].
\label{eq.AG42}
\end{equation}
The right hand side of this equation is reduced to unity in the BEC limit, ${\bar \nu}\to-\infty$, namely, the chemical potential for the $b$-Boson ($2\mu$) approaches the (renormalized) threshold energy $2{\bar \nu}$, as one expects. 
\par
The relation between $\mu$ and the binding energy of a bound state in a two-particle system can be also obtained by extending the previous discussions given in the superconductivity case.\cite{Melo,Randeria,Leggett} When we consider two Fermi atoms in a vacuum and employ relative coordinates, the scattering $t$-matrix for one Fermi atom with reduced mass $m/2$ and energy $\omega$ is given by
\begin{eqnarray}
T(\omega)
&=&
\Bigl(
-U+g_{\rm r}^2{1 \over \omega-2\nu}
\Bigr)
+
\Bigl(
-U+g_{\rm r}^2{1 \over \omega-2\nu}
\Bigr)
\sum_{\sib p}{1 \over \omega-2\varepsilon_{\sib p}}
\Bigl(
-U+g_{\rm r}^2{1 \over \omega-2\nu}
\Bigr)+\cdot\cdot\cdot
\nonumber
\\
\displaystyle
&=&
-{
U+g_{\rm r}^2{1 \over 2\nu-\omega}
\over
1-
\Bigl(
U+g_{\rm r}^2{1 \over 2\nu-\omega}
\Bigr)\sum_{\sib p}{1 \over 2\varepsilon_{\sib p}-\omega}
}.
\label{eq.AG43}
\end{eqnarray}
The ${\bf q}=0$ two-particle bound state is given by the pole of $T(\omega)$, namely
\begin{equation}
1=U'_{\rm eff}\sum_{\sib p}
{1 \over 2\varepsilon_{\sib p}-\omega},
\label{eq.AG43b}
\end{equation}
where $U'_{\rm eff}\equiv U+g_{\rm r}^2/(2\nu-\omega)$. When we renormalize the interaction in eq. (\ref{eq.AG43b}) as we have done to obtain eq. (\ref{eq.Y27}), eq. (\ref{eq.AG43b}) is reduced to
\begin{equation}
1={\bar U}'_{\rm eff}\sum_{\sib p}
\Bigl[
{1 \over 2\varepsilon_{\sib p}-\omega}
-
{1 \over 2\varepsilon_{\sib p}}
\Bigr],
\label{eq.AG44}
\end{equation}
where 
${\bar U}_{\rm eff}'\equiv{\bar U}+{\bar g}_{\rm r}^2/(2{\bar \nu}-\omega)$, in which the renormalized variables are given in eq. (\ref{eq.Y28}). This equation is the same as eq. (\ref{eq.AG43}) in the BEC limit ($\mu\ll 0$, where $\tanh(\varepsilon_{\sib p}-\mu)/2T\to 1$) when we replace $\omega\to2\mu$. Thus the energy of the bound state is related to the chemical potential by the relation $\omega=2|\mu|$, which is consistent with our earlier result in eq. (\ref{eq.Y10}).
\par
\vskip2mm
\section{Strong Coupling Effect above $T_{\rm c}$}
\vskip2mm
\subsection{BCS-BEC crossover effect on the atomic density profile}
\vskip3mm
As expected, the change of the character of the particles from the Fermion to Boson in the BCS-BEC crossover shown in Fig. 4(c) strongly affects the density profile of atoms in the trap. This is shown in Fig. 7(a). In the BCS regime, the density profile is spread out, as shown by result for $\nu=\varepsilon_{\rm F}$. In this regime, the density profile is mainly composed of Fermi atoms, and the Pauli exclusion principle between Fermi atoms effectively acts as a repulsive interaction. This pushes the atoms away from the center of the trap. As the system approaches the BEC regime ($\nu<0$), the number of stable Bosons gradually increases and thus the repulsive effect of the Pauli exclusion principle becomes less important. As expected for Bosons, the particles are now seen to cluster near the center of the trap as $\nu$ is lowered, as shown by Fig. 7(a) [see also the inset in Fig. 7(a) for $\nu=0$]. These results indicate that the easily measured atomic density profile may be a very useful signature in looking for the BCS-BEC crossover phenomenon experimentally, especially as one gets close to $T_{\rm c}$.
\par
We note that the atomic density profile shows a steep decrease near the trap center at $T_{\rm c}$ (Fig. 7(a)), but this is absent at $T=1.5T_{\rm c}$ (Fig. 7(b)). The cusp-like behavior at $T_{\rm c}$ originates from the Boson component ($N_{\rm B}$ and $N_{\rm C}$) in the density, as shown in the inset in Fig. 7(a). In the extreme BEC limit, since the gas is described by $N/2$ free Bose gas in a trap, the density of atoms is proportional to $g_{3/2}(z({\bf r}))$,\cite{Pethick} where $g_{3 \over 2}(z)\equiv\sum_{n=1}^\infty z^n/n^{3/2}$ and the fugacity is $z({\bf r})\equiv e^{(2\mu-2\nu-2V_{\rm trap}({\bf r}))/T}$. At $T_{\rm c}$, since the BEC is realized when the chemical potential of Bosons ($2\mu$) reaches the bottom of the Boson excitation spectrum ($2\nu$), we find $z({\bf r})\to 1$ at the trap center. From the well known behavior of $g_{3/2}(z)$ at $z=1$, the atomic density profile shows a finite slope at ${\bf r}=0$ at $T_{\rm c}$. In contrast, this sharp cusp at the trap center is absent for temperatures above $T_{\rm c}$, where $2\mu<2\nu$ and hence the fugacity $z({\bf r})$ is no longer close to unity.\cite{Pethick} We also show in the inset in Fig. 7(b) the density profile above $T_{\rm c}$ when the temperature is fixed as $T=0.75T_{\rm F}$. In this case, the density profile becomes more spatially diffuse as $\nu$ is increased. This is because of the increasing dominance of the Fermions and the resulting enhancement of the effect of the Pauli exclusion principle. 
\par
\vskip2mm
\subsection{Crossover from a Fermi gas to a Bose gas above $T_{\rm c}$}
\vskip2mm
It is useful to clarify when and how the stable Bosons ($N_{\rm B}^{\gamma=0}$ and $N_{\rm C}^{\gamma=0}$) at $T_{\rm c}$ shown in Fig. 4(c) disappear as the temperature is increased above $T_{\rm c}$. Figure 8(a) shows the temperature dependence of the number of particles above $T_{\rm c}$ (for $\nu=0$). The number of stable Bosons ($N_{\rm B}^{\gamma=0}$ and $N_{\rm C}^{\gamma=0}$), as well as the finite lifetime Bosons ($N_{\rm B}^{\gamma>0}$) and the scattering contribution ($N_{\rm C}^{\rm sc}$), are found to all decrease at higher temperatures. This result is also obtained when $\nu<0$. Thus only the free Fermi atoms contribute at higher temperatures well above $T_{\rm c}$. 
\par
The reason that the Fermi atom contribution dominates at high temperatures has a very simple explanation in the difference of the chemical potentials between a Fermi atom ($\mu_{\rm F}=\mu$) and a $b$-Boson ($\mu_{\rm B}=2\mu$). As the temperature is increased, the chemical potential $\mu$ must decrease so as to conserve the total number of atoms [$N$=(Fermi atoms)+2$\times$(Bose molecules)], as shown in Fig. 8(b). Thus for $T\gg T_{\rm c}$, the Boson excitation energy $E_{\sib q}^0+2\nu-2\mu$ always increases faster than the excitation energy of the Fermions $\varepsilon_{\sib p}-\mu$, simply due to the factor two in front of $\mu$ in the Bose excitation energy. As a result, the occupation of the Bose band always approaches zero for $T\gg T_{\rm c}$, even though the Bosons can be the dominant excitation at $T_{\rm c}$ in the BEC regime. 
\par
However, Fig. 8 still shows that there is a significant region above $T_{\rm c}$ where the number of long-lived Bosons is substantial. It is thus useful to define a ``crossover temperature" $T^*$ which separates the Fermi gas and the Bose gas regimes, even though the increase in the number of stable Bosons occurs continuously. We {\it define} $T^*$ as the temperature at which $N_{\rm B}^{\gamma=0}=0.05$, or when 10$\%$ of the Fermi atoms combine to form stable $b$-molecules (Fig. 9). Figure 9 shows that $T^*$ rapidly increases when $\nu$ is small or negative, for which case stable Bosons appear at $T_{\rm c}$ ($\nu/\varepsilon_{\rm F}\lesssim 0.18$, for the parameters used). Even for $\nu\simeq 0$, $T^*\sim \varepsilon_{\rm F}$ is considerably larger than $T_{\rm c}$. Since this ``Bose gas regime" lying between $T_{\rm c}$ and $T^*$ is fairly wide, one might expect to observe the formation of the stable molecules somewhat above $T_{\rm c}$ if we can decrease the threshold energy $2\nu$ enough, namely to values where the superfluid transition at $T_{\rm c}$ would be of the BEC type (see Fig. 4).
\par
Figure 10 compares the temperature dependence of the number of particles and the chemical potential (lines in the figure) for $\nu=-\varepsilon_{\rm F}$ with the case of a free ($U=g_{\rm r}=0$) Fermion-Boson mixture (filled circles and squares). In the non-interacting Fermion-Boson mixture, we only impose the constraint that the total number of atoms is conserved [$N$=(Fermi atoms)+2$\times$(Bose molecules)]. Figure 10 shows that deep in the BEC regime ($\nu=-\varepsilon_{\rm F}$), this constraint alone determines the temperature dependence of the number of particles and the chemical potential. As far as these quantities are concerned, the effect of including the interaction $g_{\rm r}$ and $U$ is clearly negligible.
\par
\vskip2mm
\section{The case of strong Feshbach coupling}
\vskip3mm
In all of the numerical results presented so far, we have considered the case of a weak Feshbach coupling described by $g_{\rm r}=0.6\varepsilon_{\rm F}$. In this regard, recent calculations have argued that strong coupling ($g_{\rm r}\gg\varepsilon_{\rm F}$) was a more realistic description of Feshbach resonances observed in ultracold gases.\cite{Holland,Kokkelman,Milstein} In this section, we investigate what arises in this strong Feshbach coupling regime, although the present treatment of fluctuations implicitly assumes a weak Feshbach coupling. We present some numerical results for $T_{\rm c}$ in the limit of a large Feshbach coupling (broad resonance) for a uniform gas. In this regard, we note that our analysis in the present paper and in Ref.\cite{Ohashi} has shown that the BCS-BEC crossover in a trap is very similar to that in a uniform gas, at least for the case of weak Feshbach coupling.
\par 
Figure 11 show the BCS-BEC crossover in the case of $g_{\rm r}=20\varepsilon_{\rm F}\gg\varepsilon_{\rm F}$. Figures 11(a) and 11(b) show that the crossover behavior already occurs around $\nu=150\varepsilon_{\rm F}$, which is much higher than $\varepsilon_{\rm F}$. Results similar to this are presented in Ref.\cite{Milstein}. Since the bottom of the Bose excitation spectrum $2\nu$ is then still much higher than the Fermion band, the Feshbach related $b$-Bosons described by $N_{\rm B}=N_{\rm B}^{\gamma=0}+N_{\rm B}^{\gamma>0}$ are almost absent in this crossover regime, as shown in Fig. 11(c). The high-energy $b$-Bosons only contribute to virtual scattering processes involving Fermions, which mediate the pairing interaction between Fermi atoms. Thus, when the Feshbach coupling $g_{\rm r}$ is very strong, the character of the BCS-BEC crossover becomes identical to that in the case of strong-coupling superconductivity as discussed by Nozi\`eres and Schmitt-Rink.\cite{Nozieres} The fact that the crossover to the BEC region occurs at such large values of $2\nu$ is easy to understand. Since the chemical potential $\mu$ is always of order of $\varepsilon_{\rm F}$ or less, at large values of $2\nu$ and $g_{\rm r}$, we have $U_{\rm eff}=U+g_{\rm r}^2/(2\nu-2\mu)\simeq g_{\rm r}^2/2\nu$. Since the strong-coupling effect should be most pronounced when $U_{\rm eff}$ becomes comparable to $\varepsilon_{\rm F}$, the BCS-BEC crossover will occur at $U_{\rm eff}\sim\varepsilon_{\rm F}$, which gives $2\nu\sim g_{\rm r}^2/\varepsilon_{\rm F}$. For $g_{\rm r}=20\varepsilon_{\rm F}$, this predicts that the crossover will occur at $\nu\sim 200\varepsilon_{\rm F}$, consistent with the numerical results in Fig. 11(a).
\par
That the case of a very broad Feshbach resonance ($g_{\rm r}\gg\varepsilon_{\rm F}$) is equivalent to the case studied in Ref.\cite{Nozieres,Randeria} is also shown clearly by the fact that stable preformed Cooper-pairs are the dominant excitation [see Fig. 11(c)] in the crossover region at $\nu\sim150\varepsilon_{\rm F}$. The Feshbach $b$-molecule component only becomes dominant when $\nu\lesssim 0$. This result is in contrast to our results for a weak Feshbach coupling $g_{\rm r}=0.6\varepsilon_{\rm F}$ as shown in Fig. 4. In that case, the $b$-molecule becomes the dominant component at the BCS-BEC crossover. 
\par
Figure 12 shows the total number of composite Bosons $N_{\rm M}\equiv N_{\rm B}^{\gamma=0}+N_{\rm B}^{\gamma>0}+N_{\rm C}^{\gamma=0}+N_{\rm C}^{\rm sc}$ as a function of $\nu$, for $g_{\rm r}=20\varepsilon_{\rm F}$. Although the individual Cooper-pair components $N_{\rm C}^{\gamma=0}$ and $N_{\rm C}^{\rm sc}$ show [see Fig. 11(c)] singular behavior in the BCS-BEC crossover regime (at $\nu\sim150\varepsilon_{\rm F}$), $N_{\rm M}$ itself increases smoothly as the threshold energy is decreased.
\par
Since the BCS-BEC crossover around $\nu=150\varepsilon_{\rm F}$ shown in Fig. 11 is intrinsically the same phenomenon as discussed by Nozi\`eres and Schmitt-Rink,\cite{Nozieres} the origin of the small peak in $T_{\rm c}$ shown in Fig. 11(a) is the analogue of the one formed also in the case of strong-coupling superconductivity.\cite{Nozieres,Melo,Randeria} However, Haussmann\cite{Haussmann} showed that the slight peak in $T_{\rm c}$ obtained at the BCS-BEC crossover in strong-coupling superconductivity disappears when one includes higher order scattering processes beyond the simple $t$-matrix approximation, such as used in the present paper. Thus, we regard the small peak $T_{\rm c}$ in Fig. 11(a) as probably an artifact of our simple $t$-matrix approximation. As in the weak Feshbach coupling case shown in Fig. 4, we expect that the maximum transition temperature $T_{\rm c}$ will again be given by the BEC expressions in eqs. (\ref{eq.3.3}) and (\ref{eq.3.2}).
\vskip2mm
\section{Summary}
\vskip2mm
To summarize, we have extended our recent work\cite{Ohashi} and investigated the effect of particle-particle fluctuations in conjunction with an atomic Feshbach resonance on the superfluid phase transition in a gas of Fermi atoms in a harmonic trap, using a simple LDA approach. The BCS-BEC crossover predicts a maximum transition temperature of $T_{\rm c}=0.518T_{\rm F}$, somewhat larger than $T_{\rm c}=0.218T_{\rm F}$ in a uniform gas.\cite{Ohashi} In the BEC regime, the excitation spectrum is dominated by the stable long-lived Feshbach-related molecules in the case of a small Feshbach coupling parameter $g_{\rm r}$. This is in contrast with results for the BCS-BEC crossover with no Feshbach resonance,\cite{Nozieres,Randeria} where the preformed Cooper-pairs are the dominant bound states. For a very broad Feshbach resonance ($g_{\rm r}\gg\varepsilon_{\rm F}$) as studied in Refs.\cite{Holland,Kokkelman,Milstein}, we find that the crossover region is also dominated by preformed Cooper-pairs.
\par
The BCS-BEC crossover is shown to strongly affect the atomic density profile in the trap, which can be measured. In the BCS regime, where Fermi atoms are dominant excitations, the spatial distribution of the atoms is spread out from the trap center due to the Pauli exclusion principle. The atoms increasingly cluster at the center of the trap as the threshold energy of $b$-Bose excitation spectrum is lowered and the the number of stable composite Bosons increases. 
\par
In the BEC regime, we show that stable molecules can exist above $T_{\rm c}$. As the temperature increases, however, the number of these Bosons decreases. We introduce a crossover temperature $T^*$ defined as the temperature above $T_{\rm c}$ at which 10$\%$ of the Fermi atoms have formed long-lived $b$-molecules. Plotting $T^*$ as a function of the threshold $2\nu$ gives a sort of ``phase diagram" in the normal phase which may be useful in experimental searches for the BCS-BEC crossover.
\par
The experimental observation of Cooper-pairs in trapped Fermi gases will be a very exciting milestone in physics. What makes this topic even more interesting is the possibility that we can observe the BCS-BEC crossover in such systems. As we have shown in this paper, this crossover involves very interesting physics and, moreover, it leads to characteristic changes in the properties of the trapped gas. The changes in the atomic density profile may provide a crucial experimental signature for the existence of Cooper pairing and superfluidity.
\par
We will discuss the BCS-BEC crossover in the superfluid region below $T_{\rm c}$ in a future paper. In particular, we will discuss how the collective modes vary as we pass through the crossover region.\cite{Ohashi2}
\par
\acknowledgements
The authors would like to acknowledge useful discussion with M. Holland and J. Milstein concerning Refs.\cite{Kokkelman,Milstein}. Y. O would like to thank Professor S. Takada for valuable discussions, as well as financial support from the Japanese Government while a visiting professor at the University of Toronto. A.G. acknowledges support from NSERC of Canada.
%
%
\newpage
\appendix
\section{Calculation of fluctuation contribution to the number of Fermi atoms}
The density of atoms associated with fluctuations given by $\delta N_{\rm FL}({\bf r})$ in eq. (\ref{eq.2.8}) can be split into two contributions [using eq. (\ref{eq.G6})],\cite{note5}
\begin{equation}
\delta N_{\rm FL}({\bf r})=
{1 \over \beta}\sum_{{\sib q},\nu_n}
e^{i\nu_n\delta}
{U \over 1-U\Pi}{\partial \Pi \over \partial\mu}
-
{1 \over \beta}\sum_{{\sib q},\nu_n}e^{i\nu_n\delta}
{\partial \over \partial\mu}
\ln
\bigl[1+g_{\rm r}^2D_0{\tilde \Pi}\bigr].
\label{eq.a.1}
\end{equation}
Since the second term involving ${\partial \over \partial\mu}\ln[1+g_{\rm r}^2D_0{\tilde \Pi}]$ goes as $\nu_n^{-2}$ in the large $\nu_n$ limit, the convergence factor $e^{i\nu_n\delta}$ has no effect. However, this convergence factor cannot be dropped in the first term in eq. (\ref{eq.a.1}), because the quantity ${U \over 1-U\Pi}{\partial \Pi \over \partial\mu}$ only goes as $\nu_n^{-1}$ for large $\nu_n$. In numerical calculations, it is difficult to sum up the Matsubara frequencies directly taking into account the convergence factor involving a infinitesimally small number $\delta$. One method to avoid this difficulty is to employ a spectral representation as carried out by Nozi\`eres and Schmitt-Rink.\cite{Nozieres} However, this method requires high numerical accuracy in executing the resulting energy integration. We used a different approach.
\par
To calculate the first term in eq. (\ref{eq.a.1}) [$\equiv\delta N_{\rm FL}^{(1)}({\bf r})$] without employing a spectral representation, we compare $\delta N_{\rm FL}^{(1)}({\bf r})$ to the expression without the convergence factor,
\begin{equation}
\delta N_{\rm FL}^{(2)}({\bf r})\equiv
{1 \over \beta}\sum_{{\sib q},\nu_n}
{U \over 1-U\Pi}{\partial \Pi \over \partial\mu}.
\label{eq.a.2}
\end{equation}
This quantity is easy to calculate numerically. When we transform the summation over the discrete Matsubara frequencies into an integral in the complex plane in the usual way, we obtain
\begin{eqnarray}
\left\{
\begin{array}{l}
\displaystyle
\delta N_{\rm FL}^{(1)}({\bf r})={1 \over 2\pi i}\sum_{\sib q}
\int_{C_0} dz 
e^{\delta z} n_{\rm B}(z)
{U \over 1-U\Pi(i\nu_n\to z)}{\partial \Pi(i\nu_n\to z) \over \partial\mu},
\\
\displaystyle
\delta N_{\rm FL}^{(2)}({\bf r})={1 \over 2\pi i}\sum_{\sib q}
\int_{C_0} dz n_{\rm B}(z)
{U \over 1-U\Pi(i\nu_n\to z)}{\partial \Pi(i\nu_n\to z) \over \partial\mu},
\end{array}
\right.
\label{eq.a.3}
\end{eqnarray}
where the path $C_0$ is shown in Fig. 13. we are allowed to add the circular contours $C_1$ and $C_2$ for the integration in $\delta N_{\rm FL}^{(1)}({\bf r})$ without changing the result, because the contributions from $C_1$ and $C_2$ vanish in the large radius limit ($R\to\infty$) due to, respectively, the factors $n_{\rm B}(z)$ and $e^{\delta z}$. The integration in $\delta N_{\rm FL}^{(1)}({\bf r})$ is thus reduced to summing up the poles inside the closed contours $C_0+C_1$ and $C_0+C_2$ shown in Fig. 13.
\par
We next show how $\delta N_{\rm FL}^{(1)}({\bf r})$ is related to $\delta N_{\rm FL}^{(2)}({\bf r})$ defined in eq. (\ref{eq.a.2}). First, we note that when dealing with $\delta N_{\rm FL}^{(2)}({\bf r})$, the integration along the path $C_2$ gives a finite contribution when we add the paths $C_1$ and $C_2$ due to the absence of the convergence factor. However, if we write the integration in $\delta N_{\rm FL}^{(2)}({\bf r})$ as
\begin{equation}
\int_{C_0}dz=\int_{C_0+C_1+C_2}dz -\int_{C2}dz,
\label{eq.a.4}
\end{equation}
we find that the first term can be calculated by summing up the contribution from the poles inside the closed paths $C_0+C_1$ and $C_0+C_2$, and the result is identical to $\delta N_{\rm FL}^{(1)}({\bf r})$. The last term in eq. (\ref{eq.a.4}) [$\equiv \delta N_{\rm FL}^{\rm corr}({\bf r})$] can be evaluated to give, in the limit $R\to\infty$,
\begin{eqnarray}
\delta N_{\rm FL}^{\rm corr}({\bf r})
&=&
-{U \over 2\pi}\sum_{{\sib p},{\sib q}}
e^{-\varepsilon_{\sib p}^2/\omega_c^2}
\int_{C_2}d\theta
\Bigl[
{\partial f(\varepsilon_{{\sib p}+{\sib q}/2}-\mu({\bf r})) \over \partial\mu}
+
{\partial f(\varepsilon_{{\sib p}-{\sib q}/2}-\mu({\bf r})) \over \partial\mu}
\Bigr]
\nonumber
\\
&=&
U
\sum_{{\sib p},{\sib q}}
e^{-\varepsilon^2_{\sib p}/\omega_c^2}
{\partial f(\varepsilon_{{\sib p}+{\sib q}/2}-\mu({\bf r})) \over \partial\mu}.
\label{eq.a.5}
\end{eqnarray}
Here, $e^{-\varepsilon^2_{\sib p}/\omega_c^2}$ is the Gaussian cutoff introduced in Section IV, and we have used eq. (\ref{eq.2.4}) and changed the integration variable using $z=Re^{i\theta}$. Thus we have shown that $\delta N_{\rm FL}^{(1)}({\bf r})$ involving the convergence factor is related to $N_{\rm FL}^{(2)}({\bf r})$ without the convergence factor by
\begin{equation}
\delta N_{\rm FL}^{(1)}({\bf r})=\delta N_{\rm FL}^{(2)}({\bf r})-\delta N_{\rm FL}^{\rm corr}({\bf r}).
\label{eq.a.6}
\end{equation}
Using this expression, we can calculate $\delta N_{\rm FL}^{(1)}({\bf r})$ numerically by evaluating the simpler expression $\delta N_{\rm FL}^{(2)}({\bf r})$.
\par
The essence of this prescription may be easily understood from the following example: Consider the two quantities:
\begin{eqnarray}
\left\{
\begin{array}{l}
\displaystyle
N_{\rm B}^{(1)}({\bf q})=-\sum_{\nu_n} e^{i\nu_n\delta}{1 \over i\nu_n-E_{\sib q}},
\\
\displaystyle
N_{\rm B}^{(2)}({\bf q})=-\sum_{\nu_n}{1 \over i\nu_n-E_{\sib q}}.
\end{array}
\right.
\label{eq.a.7}
\end{eqnarray}
In this case, the frequency sum in $N_{\rm B}^{(1)}({\bf q})$ can be easily done analytically and reduced to the Bose distribution function $n_{\rm B}(E_{\sib q})$. The sum in $N_{\rm B}^{(2)}$ gives
\begin{equation}
N_{\rm B}^{(2)}({\bf q})=\sum_{\nu_n}{E_{\sib q} \over \nu_n^2+E^2_{\sib q}}
={1 \over 2}{\rm coth}{\beta E_{\sib q} \over 2}
.
\end{equation}
The contribution from $C_2$ in the last term in eq. (\ref{eq.a.4}) is given by
\begin{equation}
\delta N_{\rm B}^{\rm corr}\equiv-{1 \over 2\pi i}\int_{C_2:R\to\infty} dz
{1 \over z-E_{\sib q}}={1 \over 2}.
\end{equation}
Thus we can easily verify that $N_{\rm B}^{(1)}=N_{\rm B}^{(2)}-\delta N_{\rm B}^{\rm corr}$, as in eq. (\ref{eq.a.6}).
%

%
\begin{figure}
\caption{Generalized $t$-matrix approximation for the particle-particle scattering vertex function $\Gamma(p_+,-p_-,p'_+,-p'_-)$. The solid and dashed lines represent the one-particle thermal Fermion Green function $G_0$ and Boson Green function $D_0$, respectively. The first line includes the ladder processes associated with the direct pairing interaction $-U$. The second line includes the effect of the Feshbach resonance with the coupling $g_{\rm r}$. The shaded bubble in the second line includes the ladder diagrams shown in the third line.
}
\label{Fig1} 
\end{figure}

\begin{figure}
\caption{Fluctuation contributions to the thermodynamic potential $\Omega$. (a) Ladder diagrams associated with the direct pairing interaction $-U$. (b) Diagrams arising in the presence of the Feshbach resonance. The shaded bubble is the same as in Fig. 1.
}
\label{Fig2} 
\end{figure}

\begin{figure}
\caption{(a) Renormalization of the non-resonant pairing interaction $-U$, as given in eq. (4.4). The solid line with solid circle means the Fermion Green function in the high-energy region $[\omega_c,E_c]$. (b) Renormalization of the Feshbach resonance coupling $g_{\rm r}$, which is described by the three point vertex correction. (c) Renormalization of the threshold energy $2\nu$ associated with the self-energy correction.
}
\label{Fig3} 
\end{figure}

\begin{figure}
\caption{BCS-BEC crossover in a trapped Fermi gas, for the parameters $U/\varepsilon_{\rm F}=0.3$ and $g_{\rm r}/\varepsilon_{\rm F}=0.6$. (a) The superfluid phase transition temperature $T_{\rm c}$ as a function of $\nu$. In this figure, BCS is the result neglecting particle-particle fluctuations while BEC is the value for $T_{\rm c}$ for a gas of $N/2$ Bosons of mass $M=2m$. (b) The chemical potential $\mu$ at $T_{\rm c}$ as a function of $\nu$. The dashed line ($2\mu=2\nu$) shows the chemical potential for a trapped gas of $N/2$ $b$-Bosons. (c) Change in the character of particles through the BCS-BEC crossover region. We note that scattering contribution $N_{\rm C}^{\rm sc}$ becomes negative below $\nu/\varepsilon_{\rm F}\simeq 0.18$, where the stable Feshbach molecule ($N_{\rm B}^{\gamma=0}$) and preformed Cooper-pair ($N_{\rm C}^{\gamma=0}$) components first appear. The sum $N_{\rm C}^{\gamma=0}+N_{\rm C}^{\rm sc}$ is positive.
}
\label{Fig4} 
\end{figure}

\begin{figure}
\caption{The spectral density $\rho_{\rm B}$ of the renormalized Bose Green function ${\tilde D}$, for $T=T_{\rm c}$ and ${\bf r}=0$ and $q=p_F$, where $p_{\rm F}$ is the Fermi momentum for a free gas of $N/2$ spin up Fermions. In each panel, the inset shows the frequency dependence of the damping $\gamma(\omega)$.
}
\label{Fig5} 
\end{figure}

\begin{figure}
\caption{The number of Bosons $N_{\rm M}$ [see eqs. (3.16) and (3.17)], as a function of the threshold energy $\nu$. The number of Fermi atoms $N_{\rm F}^0$ is also shown.
}
\label{Fig6} 
\end{figure}

\begin{figure}
\caption{Atomic density profile at $T_{\rm c}$ as a function of position in the trap in the BCS-BEC crossover. (a) $T=T_{\rm c}$. (b) $T=1.5T_{\rm c}$. Inset in Fig. 7(a): Density profiles of $N_{\rm F}^0({\bf r})$, $N_{\rm B}^{\gamma=0}({\bf r})$ and $N_{\rm C}^{\gamma=0}({\bf r})$ at $T=T_{\rm c}$ in the case of $\nu=0$. Inset in Fig. 7(b): Density profile above $T_{\rm c}$ when the temperature is fixed as $T=0.75T_{\rm F}$. These results are for $U/\varepsilon_{\rm F}=0.3$ and $g_{\rm r}/\varepsilon_{\rm F}=0.6$. The number density of atoms $N({\bf r})$ is related to the total number of atoms $N$ as $N=\int d{\bf r}N({\bf r})$. One Boson counts as two Fermi atoms in this figure. We note that different temperatures are used for the three cases ($\nu/\varepsilon_{\rm F}=\pm1, 0$) in Fig. 7(b). Since $T_{\rm c}$ increases as $\nu$ is lowered [see Fig. 4(a)], the temperature for $\nu=\varepsilon_{\rm F}$ in Fig. 7(b) is the lowest among the three cases plotted. As a result, the spatial diffusion of atoms in the density profile originating from the increase of the (averaged) kinetic energy of atoms due to a finite temperature is weakest in the case of $\nu=\varepsilon_{\rm F}$ and hence the density profile at the trap center is largest.
}
\label{Fig7} 
\end{figure}

\begin{figure}
\caption{
(a) The number of particles above $T_{\rm c}$ when the threshold energy is zero, for $U/\varepsilon_{\rm F}=0.3$ and $g_{\rm r}/\varepsilon_{\rm F}=0.6$. (b) Temperature dependence of the chemical potential $\mu$ above $T_{\rm c}$.
}
\label{Fig8} 
\end{figure}

\begin{figure}
\caption{Schematic phase diagram for Bose molecule phase. For a given value of $\nu$, the temperature $T^*(\nu)$ is defined where the number of stable Feshbach molecules $N_{\rm B}^{\gamma=0}$ equals $0.05N$. These results are for $U/\varepsilon_{\rm F}=0.3$ and $g_{\rm r}/\varepsilon_{\rm F}=0.6$. Above $T^*(\nu)$, the system can be regarded as a gas of Fermi atoms, while below this temperature, one has an increasing fraction of pairing into stable molecules. The number of stable Feshbach molecules and the preformed Cooper-pairs at $T_{\rm c}$ are also shown.
}
\label{Fig9} 
\end{figure}

\begin{figure}
\caption{(a) The number of Fermi atoms $N_{\rm F}^0$ and the stable Feshbach molecule component $N_{\rm B}^{\gamma=0}$ for temperatures above $T_{\rm c}$, for $\nu=-\varepsilon_{\rm F}$. (b) The chemical potential $\mu$ above $T_{\rm c}$. The lines show the results for $U/\varepsilon_{\rm F}=0.3$ and $g_{\rm r}/\varepsilon_{\rm F}=0.6$. The circles and squares are the results for a non-interacting Boson-Fermion gas ($U=g_{\rm r}=0$) but with the constraint that the total number of the particles $N=$[Fermi atoms+2$\times$Bose molecules] is fixed.
}
\label{Fig10} 
\end{figure}

\begin{figure}
\caption{BCS-BEC crossover in a trapped Fermi gas in the case of a strong Feshbach coupling $g_{\rm r}/\varepsilon_{\rm F}=20$. We take $U/\varepsilon_{\rm F}=0.3$. (a) $T_{\rm c}$ as a function of $\nu$. (b) The chemical potential $\mu$ at $T_{\rm c}$. (c) Change of the character of particles through the BCS-BEC crossover. The number of stable Cooper-pairs $N_{\rm C}^{\gamma=0}$ exceeds $N/2~(=0.5N)$ around the sharp peak shown in panel (c). However this is canceled by the negative value of the scattering contribution $N_{\rm C}^{\rm sc}$. The total number of Bosons is always smaller than $N/2$ (see the graph of $N_{\rm M}$ in Fig. 12). }
\label{Fig11} 
\end{figure}

\begin{figure}
\caption{Change of the number of Bosons $N_{\rm M}$ in the BCS-BEC crossover in the case of strong Feshbach coupling, taking $g_{\rm r}/\varepsilon_{\rm F}=20$ and $U/\varepsilon_{\rm F}=0.3$. In this figure, the contribution from $N_{\rm B}^{\gamma>0}$ is negligible. Compare with the analogous results for weak Feshbach coupling in Fig. 6.
}
\label{Fig12} 
\end{figure}

\begin{figure}
\caption{Path for the complex integration discussed in the Appendix. The dots show the imaginary discrete Bose Matsubara frequencies $i\nu_n$.
}
\label{Fig13} 
\end{figure}

%
%
%

\begin{references}
\bibitem{Stoof} For a review, see H. T. C. Stoof and M. Houbiers, {\it Bose-Einstein Condensation in Atomic Gases}, ed. M. Inguscio, S. Stringari and C. E. Wieman (IOS press, Amsterdam, 1999), p.537 (1999).
\bibitem{Jin} B. DeMarco and D. S. Jin, Science {\bf 285}, 1703 (1999).
\bibitem{Andrew} A. G. Truscott, K. E. Strecker, W. I. McAlexander, G. B. Partridge and R. G. Hulet, Science {\bf 291}, 2570 (2001).
\bibitem{Salomon} F. Schreck, L. Khaykovich, K. L. Corwin, G. Ferrari, T. Bourdel, J. Cubizolles and C. Salomon, Phys. Rev. Lett. {\bf 87}, 080403 (2001).
\bibitem{Granade} S. R. Granade, M. E. Gehm, K. M. O'Hara and J. E. Thomas, Phys. Rev. Lett. {\bf 88}, 120405 (2002).
\bibitem{Loftus} T. Loftus, C. A. Regel, C. Ticknor, J. L. Bohn and D. S. Jin, Phys. Rev. Lett. {\bf 88}, 173201 (2002).
\bibitem{Timmermans1} E. Timmermans, P. Tommasini, M. Hussein and A. K. Kerman, Phys. Rep. {\bf 315}, 199 (1999).
\bibitem{Timmermans2} E. Timmermans, K. Furuya, P. W. Milonni and A. K. Kerman, Phys. Lett. A {\bf 285}, 228 (2001).
\bibitem{Holland} M. Holland, S. J. J. M. F. Kokkelmans, M. L. Chiofalo and R. Walser, Phys. Rev. Lett. {\bf 87}, 120406 (2001).
\bibitem{Chiofalo} M. L. Chiofalo, S. J. J. M. F. Kokkelmans, 
J. N. Milstein and M. J. Holland, Phys. Rev. Lett. {\bf 88}, 090402 (2002).
\bibitem{Ohashi} Y. Ohashi and A. Griffin, Phys. Rev. Lett. {\bf 89}, 130402 (2002).
\bibitem{Nozieres} P. Nozi\`eres and S. Schmitt-Rink, J. Low. Temp. Phys. {\bf 59}, 195 (1985).
\bibitem{Tokumitsu} A. Tokumitsu, K. Miyake and K. Yamada, Phys. Rev. B {\bf 47}, 11988 (1993).
\bibitem{Melo} C. A. R. S\'a de Melo, R. Randeria and J. R. Engelbrecht, Phys. Rev. Lett. {\bf 71}, 3202 (1993).
\bibitem{Randeria} For a review, see M. Randeria, in {\it Bose-Einstein Condensation}, ed. A. Griffin, D. W. Snoke and S. Stringari (Cambridge, N.Y. 1995) p.355.
\bibitem{Kokkelman} S. J. J. M. F. Kokkelmans, J. N. Milstein, M. L. Chiofalo, R. Walser and M. J. Holland, Phys. Rev. A {\bf 65}, 053617 (2002).
\bibitem{Milstein} J. N. Milstein, S. J. J. M. F. Kokkelmans and M. J. Holland, cond-mat/0204334v2.
\bibitem{Lee} R. Friedberg and T. D. Lee, Phys. Rev. B. {\bf 40}, 6745 (1989).
\bibitem{Ranninger1} J. Ranninger, in {\it Bose-Einstein Condensation}, ed. A. Griffin, D. W. Snoke and S. Stringari (Cambridge, N.Y. 1995), p.393.
\bibitem{Ranninger2} T. Kostyrko and J. Ranninger, Phys. Rev. B {\bf 54}, 13105 (1996).
\bibitem{Molmer} K. Molmer, Phys. Rev. Lett. {\bf 80}, 1804 (1998).
\bibitem{Minguzzi} A. Minguzzi, S. Conti and M. Tosi, J. Phys. Condens. Matt. {\bf 9}, L33 (1997).
\bibitem{Fulde} P. Fulde and R. Ferrell, Phys. Rev. {\bf 135}, A550 (1964).
\bibitem{Larkin} A. Larkin and Y. Ovchinnikov, Sov. Phys. JETP {\bf 20}, 762 (1965).
\bibitem{Takada} S. Takada and T. Izuyama, Prog. Theor. Phys. {\bf 41}, 635 (1969).
\bibitem{OhashiFF} Y. Ohashi, J. Phys. Soc. Jpn., in press.
\bibitem{Baym} L. P. Kadanoff and G. Baym, {\it Quantum Statistical Mechanics} (W. A. Benjamin, Inc., N. Y. 1962), Chapter 13.
\bibitem{Kadanoff} L. P. Kadanoff and P. C. Martin, Phys. Rev. {\bf 124}, 670 (1961).
\bibitem{OhashiT} Y. Ohashi and S. Takada, J. Phys. Soc. Jpn. {\bf 66}, 2437 (1997).
\bibitem{Takada2} K. Y. M. Wong and S. Takada, Phys. Rev. B {\bf 37}, 5644 (1988).
\bibitem{note} A position-dependent $T_{\rm c}$ is an artifact of using the LDA, where the effect of inhomogeneity of the system enters into the theory only through the local density.
\bibitem{Mermin} See for example, A. W. Ashcroft and N. D. Mermin, {\it Solid State Physics} (Holt, Rinehart, and Winston, N.Y. 1976), Chapter 2.
\bibitem{AGD} See for example, A. A. Abrikosov, L. P. Gor'kov and I. E. Dzyaloshinski, {\it Methods of Quantum Field Theory in Statistical Mechanics} (Dover, N. Y. 1963), Chapter 3.
\bibitem{Gorkov} L. P. Gor'kov and T. K. Melik-Barkhudarov, Sov. Phys. JETP {\bf 13}, 1018 (1961). 
\bibitem{Combescot} R. Combescot, Phys. Rev. Lett. {\bf 83}, 3766 (1999), see also C. J. Pethick and H. Smith, {\it Bose-Einstein Condensation in Dilute Bose Gases} (Cambridge, 2002), Chapter 14.
\bibitem{noteY} In the case of the Gaussian cutoff used in this paper, we should insert $F_{\rm cut}=e^{-\varepsilon^2_{\sib p}/\omega_c^2}$ in the summation in the low energy part and the factor $(1-F_{\rm cut})$ in the high energy part, taking the summation over $[0,E_D]$ in both contributions. This effectively gives the same result discussed in the text.
\bibitem{Pethick2} See for example, C. J. Pethick and H. Smith, {\it Bose-Einstein Condensation in Dilute Bose Gases} (Cambridge, 2002), Chapter 14.
\bibitem{Pethick} See for example, C. J. Pethick and H. Smith, {\it Bose-Einstein Condensation in Dilute Bose Gases} (Cambridge, 2002), Chapter 2.
\bibitem{Leggett} A. J. Leggett, in {\it Modern Trends in the Theory of Condensed Matter}, ed. by A. Pekalski and J. Przystawa (Springer Verlag, 1980), p.14.
\bibitem{note33} Equation (\ref{eq.AG42}) shows that $\mu>{\bar \nu}$, which is different from the result in Fig. 4(b) showing $\mu<\nu$. This is, however, simply due to the different renormalization we use in eq. (\ref{eq.AG42}). 
\bibitem{Haussmann} R. Haussmann, Phys. Rev. B {\bf 49}, 12975 (1994).
\bibitem{Ohashi2} Y. Ohashi and A. Griffin, in preparation.
\bibitem{note5} We also add a cutoff $e^{-\varepsilon_{\sib p}^2/\omega_c^2}$ in the summation in terms of ${\bf p}$, as discussed in Section IV. Thus the momentum summation in $\delta N_{\rm FL}({\bf r})$ always converges.
\end{references}
\end{document}